\documentclass{article}

\usepackage{arxiv}

\usepackage[utf8]{inputenc} % allow utf-8 input
\usepackage[T1]{fontenc}    % use 8-bit T1 fonts

\usepackage{booktabs}       % professional-quality tables
\usepackage{nicefrac}       % compact symbols for 1/2, etc.
\usepackage{microtype}      % microtypography
\usepackage{lipsum}		% Can be removed after putting your text content
\usepackage{doi}

\usepackage{hyperref, url}
\usepackage{amsfonts, amsmath, amssymb, parskip, bbm}
\usepackage{xcolor, graphicx, tikz-cd, caption, subcaption}
\usepackage{algorithmic, algorithm}
\usepackage[verbatim]{minted}
\usepackage{verbatim}
\usepackage{mathrsfs}

\usepackage{amsthm}  % For proof

\newtheorem{remark}{Remark}
\usepackage[most]{tcolorbox}

\definecolor{ink}{HTML}{17233C}
\definecolor{greenlight}{HTML}{EEF6F1}
\newtcolorbox{algorithmidea}[1]{
  enhanced,
  breakable,
  colback=greenlight,
  colframe=greenlight,
  boxrule=0pt,
  arc=1mm,
  left=8pt,right=8pt,top=7pt,bottom=7pt,
  before skip=8pt,after skip=8pt,
  title={Algorithmic construction: #1},
  fonttitle=\bfseries\small,
  fontupper=\small,
  coltitle=ink,
  attach title to upper={\par\smallskip}
}

\title{From Continuous Dynamics to Practical Gradient-Based Samplers}

\date{} 					% Or removing it

\author{
  \hspace{1mm}James Chok \\
  School of Mathematics and Maxwell Institute for Mathematical Sciences, \\
  The University of Edinburgh, Edinburgh, EH9 3FD, United Kingdom\\
  \texttt{james.chok@ed.ac.uk} \\
}

\renewcommand{\shorttitle}{Continuous Dynamics to Practical Gradient-Based Samplers}

\hypersetup{
pdftitle={Continuous Dynamics to Practical Gradient-Based Samplers},
pdfsubject={stats.CO},
pdfauthor={James Chok},
pdfkeywords={Langevin, Hamiltonian, MCMC, Preconditioning},
}

\begin{document}
\maketitle

\begin{abstract}
	Gradient-based Markov chain Monte Carlo methods are often introduced as a catalog of algorithms: Hamiltonian Monte Carlo (HMC), the Metropolis-adjusted Langevin algorithm (MALA), the No-U-Turn Sampler (NUTS), and several underdamped variants. This presentation obscures the common structure of the methods and, more importantly, the reasons why a sampler that is correct in principle may be ineffective in practice. We develop a unified account, beginning with exact continuous-time dynamics that represent idealized sampling methods and for which Metropolis adjustments are not required. Numerical discretization makes the dynamics computationally feasible but introduces bias. Metropolis adjustment removes the asymptotic bias by converting numerical errors into rejection, leading to HMC, MALA, NUTS, and the Metropolis-adjusted kinetic Langevin algorithm (MAKLA).

    The second half of the paper presents geometric design choices that determine practical performance, namely, although MAKLA and NUTS have nice theoretical properties, their sampling efficiency may be slow in practice. Importantly, a fixed mass matrix can whiten globally anisotropic targets, often fixing sampling inefficiency in Bayesian posteriors with large data. Whereas hierarchical posteriors introduce their own problem, causing state-dependent variation in the Hessian (e.g., Neal's funnel). We explain how a randomized step size can be used effectively to sample from such a distribution. The resulting paper is both a tutorial on the mechanics of gradient-based sampling and a set of practical recipes to improve sampler performance. 
\end{abstract}

% keywords can be removed

\section{Introduction}
Let $\pi(x)$ be a probability density on $\mathbb{R}^d$ known up to a normalizing constant, assumed to take the form
\begin{equation}
    \pi(x)\propto e^{-U(x)}.\notag
\end{equation}
In Bayesian inference, this typically takes the form
\begin{equation}
    \pi(x)\propto p(y|x)p(x),\qquad U(x)=-\log \big[p(y|x)p(x)\big],\notag
\end{equation}
where $p(y|x)$ is the likelihood, and $p(x)$ is the prior. When direct independent samples are unavailable, one typically builds a Markov chain Monte Carlo to construct a transition kernel whose invariant distribution is $\pi$. While this statement is mathematically precise, one often forgets the computational perspective: how fast does the chain approach equilibrium, and whether the numerical method remains stable across relevant posterior distributions?

Gradient-based samplers exploit the score $\nabla\log\pi(x)=-\nabla U(x)$. With the advent of auto-differentiation (e.g., \texttt{PyTorch}, \texttt{JAX}, and \texttt{STAN}), which allows us to compute these gradients without explicit formulas, gradient-based samplers have proven to be a powerful technique for yielding fast-converging Markov chains.

The central organizing principle of this paper is as follows
\begin{itemize}
    \item Continuous dynamics: These represent idealized samplers, utilizing explicit formulas for the evolution of the dynamics, which target $\pi(x)$ as the invariant distribution.
    \item Discretizations: Practical applications do not have explicit formulas for how the dynamics evolve. As such, one must resort to their discretized variants, and no longer target $\pi$ as the invariant distribution. Often, these target distributions are $\varepsilon$-close to $\pi$, where the distance $\varepsilon$ can be reduced by decreasing the step-size.
    \item Metropolis correction: Adding a Metropolis--Hastings filter (an accept/reject stage) restores exact invariance, allowing the chain to target $\pi$ as the invariant distribution.
    \item Geometric adaptation: While the performance of gradient-based samplers is excellent, they can often be slow due to the geometry of the distribution $\pi$. Practical implementations must also account for these geometric issues in order to build an efficient sampler.
\end{itemize}

This perspective separates four errors that are frequently conflated
\begin{enumerate}
    \item Transient error: The chain has not yet approached its stationary distribution
    \item Monte Carlo error: a finite number of correlated stationary draws is used to estimate an expectation
    \item Discretization bias: The numerical Markov chain (before a Metropolis-adjustment) has a stationary distribution different from $\pi$.
    \item Numerical instability: The integrator fails to resolve the dynamics, leading to exploding trajectories or vanishing acceptance probabilities.
\end{enumerate}
Exact continuous dynamics eliminate the third and fourth errors, but not necessarily the first two. An exact transition can preserve $\pi$ while remaining highly correlated. Conversely, a biased unadjusted chain may yield useful predictions quickly if the inferential goal is tolerant of a small, well-diagnosed perturbation. The relevant question is therefore not simply whether an algorithm is exact, but what errors one is willing to accept given the computation budget.\footnote{All code can be found on our GitHub \url{https://github.com/infamoussoap/OnSamplingMethods}.}

\section{Continuous-time Gradient-based Samplers}
Before discussing practical implementations of gradient-based samplers, we first discuss the continuous-time samplers they are based on. The Metropolis adjustments can be seen as a correction to the bias introduced by time discretization.

\subsection{Exact Hamiltonian Monte Carlo}
Define the Hamiltonian
\begin{equation}
    H(x,p)=U(x)+K(p),\qquad K(p)=\frac{1}{2}p^\top M^{-1}p,\notag
\end{equation}
where $x$ is the position, and $p$ is the momentum. Hamilton's equations are
\begin{equation}\label{eq:hamiltons_equations}
    \dot{x}=M^{-1}p,\qquad \dot{p}=-\nabla U(x).
\end{equation}
Let $\Phi_t(x,p)$ denote the exact flow, i.e., given an initial condition $(x_0,p_0)$, $\Phi_t(x_0,p_0)=(x_t,p_t)$ evolves the initial condition under \eqref{eq:hamiltons_equations} for time $t$.

\begin{algorithmidea}{Idealized Hamiltonian Monte Carlo (IHMC)}
Let $x_i$ be the current state of the Markov chain
\begin{enumerate}
    \item Draw a fresh momentum: $p_{i}\sim\mathcal{N}(0,M)$
    \item Evolve along the Hamiltonian trajectory for integration time $T$, $(x_i,p_i)\mapsto \Phi_T(x_i,p_i)=(x_{i+1},p_{i+1})$
    \item Set the new state as $x_{i+1}$
\end{enumerate}
If one also wishes to store the momentum, the invariant distribution in the extended state space is $\widetilde{\pi}(x,p)\propto\exp(-H(x,p))$. This is not needed due to the full momentum refresh at each step. 
\end{algorithmidea}  

In the idealized HMC \cite{Duane1987}, no accept/reject step is needed. Therefore, HMC crucially depends on the momentum refresh and the integration time. Without momentum refreshment, Hamiltonian evolution preserves the initial energy level and does not generally converge to the full target distribution. If $T$ is too short, the position barely changes and becomes dominated by noise. If $T$ is too long, it can return close to its starting point. Exactness, therefore, removes integration bias but not trajectory-selection problems \cite{Neal2011, BouRabee2017, hoffman2014no}.

\subsection{Overdamped Langevin dynamics}
For a fixed symmetric positive-definite matrix $C$, the preconditioned overdamped Langevin (also known simply as Langevin) diffusion is
\begin{equation}\label{eq:overdamped_langevin}
    dX_t = -C\nabla U(X_t)dt + \sqrt{2C}dW_t,
\end{equation}
where $W_t$ is the standard Brownian motion in $\mathbb{R}^d$. Under mild conditions of $\pi$, \eqref{eq:overdamped_langevin} has $\pi$ as its invariant distribution \cite{Pavliotis2014}. 

In contrast to idealized HMC, which yields a discrete Markov chain, overdamped Langevin dynamics define a continuous-time Markov chain that converges to $\pi$. However, over a short time interval $dt< 1$, deterministic travel is of order $dt$ while the stochastic travel is of order $\sqrt{dt}$ ($\geq dt$, since $dt<1$). As such, the path is dominated by stochastic noise, and long-distance movement may require many local steps.

\subsection{Underdamped Langevin Dynamics}
Underdamped Langevin (also known as kinetic Langevin) dynamics introduce momentum and combine Hamiltonian dynamics with random refreshment:
\begin{align}
    \begin{split}
    dX_t&=M^{-1}P_tdt\\
    dP_t&=-\nabla U(X_t)dt - \gamma P_t dt +\sqrt{2\gamma M} dW_t.
    \end{split}
\end{align}
Much the same as HMC, the invariant density is $\widetilde{\pi}(x,p)\propto\exp(-H(x,p))$ \cite{Pavliotis2014}. 

The friction $\gamma>0$ controls persistence. Small $\gamma$ yields trajectories similar to HMC, and in the large-friction limit, after an appropriate rescaling of time, the position process approaches overdamped Langevin dynamics. Thus, underdamped Langevin is not merely another algorithm: it is a bridge between deterministic long-trajectory HMC and local diffusion Langevin sampling.\footnote{This connection has long been known in statistical mechanics, in which underdamped Langevin can be derived as a limit from HMC, and overdamped Langevin is derived as a limit of underdamped Langevin \cite{Horowitz1991, Girolami2011, durand2023}.}

\subsection{The Gaussian Laboratory}\label{sec:gaussian_lab}
The above idealized samplers can seem, in some sense, abstract. To make them more concrete, consider the standard Gaussian target
\begin{equation}
    \pi(x)=\mathcal{N}(0,I),\qquad U(x)=\frac{1}{2}\|x\|^2.\notag
\end{equation}
This was chosen because all three continuous dynamics are exactly solvable and reveals the difference in convergence without numerical complications. Python code that uses a single exact transition for Exact HMC, overdamped and underdamped Langevin is shown in Appendix~\ref{app:gaussian_lab_python_code}.

\paragraph{Exact HMC.} With $M=I$, Hamilton's equations reduce to $\dot{x}=p$, $\dot{p}=-x$, with exact solution
\begin{equation}\label{eq:gaussian_hmc_flow}
    \Phi_t(x_i,p_i)=\begin{pmatrix}
        x_{i+1}\\ p_{i+1}
    \end{pmatrix} = \begin{pmatrix}
        x_i\cos t + p_i\sin t \\ p_i\cos t-x_i\sin t
    \end{pmatrix}.\notag
\end{equation}
\begin{algorithmidea}{Identity Gaussian Exact HMC}
    Let $x_i$ be the current state of the chain.
    \begin{enumerate}
        \item Draw a fresh momentum $p_i\sim\mathcal{N}(0, I)$.
        \item Integrate along the Hamiltonian for time $T$, $(x_{i+1},p_{i+1})=\Phi_T(x_i,p_i)$, so that
        \begin{equation}
            x_{i+1}=x_i\cos T + p_i\sin T.\notag
        \end{equation}
        \item Set $x_{i+1}$ as the next state of the chain.
    \end{enumerate}
\end{algorithmidea}

\begin{figure}
    \centering
    \includegraphics[width=0.9\linewidth]{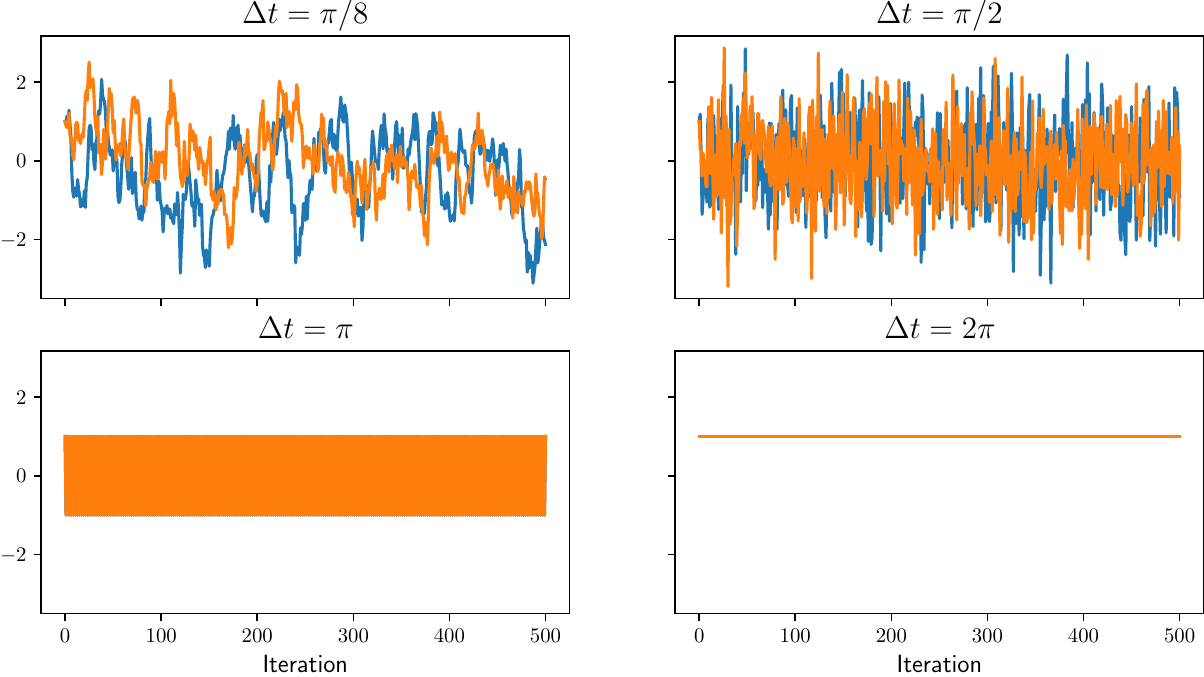}
    \caption{Exact HMC with different integration times for dimension $d=2$. From the top left, moving clockwise, HMC transitions from slow mixing to independence, then to oscillations, and finally to non-movement. The blue and orange lines are the trace plots for the first and second dimensions, respectively.}\vspace{-1em}
    \label{fig:exact_hmc}
\end{figure}

As seen in Figure~\ref{fig:exact_hmc}, the problem with HMC comes with the integration time. For an initial condition $(x_i,p_i)$, where $p_i\sim\mathcal{N}(0,I)$ is drawn at the beginning of every iteration,
\begin{itemize}
    \item For $T=\pi/8$, the exact Hamiltonian flow gives $x_{i+1}\approx 0.92 x_i + 0.38 p_i$. As such, the next state of the Markov chain $x_{i+1}$ retains much of the memory of the previous state $x_i$. This results in slow mixing chains.
    \item For $T=\pi/2$, the exact Hamiltonian flow gives $x_{i+1}=p_i$. Recalling that $p_i$ is drawn independently at every step, $x_{i+1}=p_i\sim\mathcal{N}(0,I)$, and thus is independent of the previous state of the Markov chain $x_i$. This is an exceptional quarter-period transition that yields an exact independent sampler.
    \item For $T=\pi$, the exact Hamiltonian flow gives $x_{i+1}=-x_i$. Consequently, the chain alternates deterministically between $x_0$ and $-x_0$, retaining complete dependence on its initial condition and therefore failing to converge to the target distribution. 
    \item For $T=2\pi$, the exact Hamiltonian flow gives $x_{i+1}=x_i$. The chain therefore remains fixed to its initial position $x_0$, retains complete dependence on its initial condition, and fails to converge to the target distribution.
\end{itemize}
The use of exact HMC and any of its discretized variants, therefore, must be careful in how one chooses the integration length. If too long or too short, the chain may oscillate or move slowly throughout the state space. The No-U-Turn Sampler \cite{hoffman2014no} is a derivative of HMC that has largely solved the integration-length problem.

\paragraph{Exact Overdamped Langevin:} For $M=I$, the Exact Overdamped Langevin has an iterative form
\begin{algorithmidea}{Identity Gaussian Exact Overdamped Langevin}
    Let $X_n$ be the current state of the chain, and $h$ be a fixed step size.
    \begin{enumerate}
        \item Set the new state as
        \begin{equation}
            X_{n+1} = e^{-h}X_n + \sqrt{1-e^{-2h}}\xi_n,\qquad \xi_n\sim\mathcal{N}(0,I). \notag
        \end{equation}
    \end{enumerate}
\end{algorithmidea}
For $n$ iterations, with a fixed step size, let $t=hn$ be the total integration time. Then we can also write the exact transition explicitly, 
\begin{equation}
    X_t = e^{-t}X_0+\sqrt{1-e^{-2t}}\xi,\qquad \xi\sim\mathcal{N}(0,I).\notag
\end{equation}
From a fixed initial condition $X_0\in\mathbb{R}^d$, we see that
\begin{equation}
    X_t\sim\mathcal{N}(e^{-t}X_0,(1-e^{-2t})I),\notag
\end{equation}
which approaches the target only as $t\to\infty$. In contrast to the quarter-period HMC, the overdamped Langevin only approaches the target as $t\to\infty$. As such, exact integration eliminates discretization bias but not finite-time correlation.

\paragraph{Exact Underdamped Langevin:} We omit the exact underdamped Langevin formulation, due to the relatively more complex equations. Exact formulas can be obtained and are shown in Appendix~\ref{sec:id_gaus_underdamped_langevin}.

\section{Unadjusted Discretizations}
Outside quadratic, or otherwise specially-structured targets, neither Hamiltonian nor Langevin dynamics can be solved analytically. A computer must apply a discrete map with step size $h$. The numerical chain must balance two problems:
\begin{enumerate}
    \item A small $h$ imposed by curvature to ensure numerically stable integrated trajectories
    \item Total integration time required for useful exploration of the state space.
\end{enumerate}
A smaller $h$ generally improves local accuracy and stability but requires more steps (at the cost of additional computational runtime) to cover the same integration time. Without a Metropolis correction, since reducing $h$ improves the accuracy of the numerical trajectory compared to the true trajectory, it can also reduce the bias introduced by the numerical scheme.

\subsection{Unadjusted Langevin Algorithm}
We begin with the overdamped Langevin equation, as its simple form yields explicit formulas that directly illustrate the problems with unadjusted methods. 
\begin{algorithmidea}{Unadjusted Langevin Algorithm (ULA)}
    Let $X_n$ be the current state of the chain.
    \begin{itemize}
        \item Propose the new point according to the Euler--Maruyama discretization
        \begin{equation}\label{eq:ula}
            X_{n+1}=X_n-hC\nabla U(X_n)+\sqrt{2hC}\xi_n,\qquad \xi_n\sim\mathcal{N}(0,I).
        \end{equation}
        \item Set the new state as $X_{n+1}$.
    \end{itemize}
\end{algorithmidea}
Under suitable conditions on $U(x)$ and a fixed $h$, the chain has an invariant law $\pi_h\neq \pi$.\footnote{However, the Wasserstein-2 distance $\mathcal{W}_2^2(\pi,\pi_h)$ decreases when $h$ decreases.} The one-dimensional Gaussian is especially transparent. Let $d=1$, $U(x)=x^2/2$, and $M=1$. Then
\begin{equation}
    X_{n+1}=(1-h)X_{n}+\sqrt{2h}\xi_n.\notag
\end{equation}
Stability therefore requires $|1-h|<1$, equivalently, $0<h<2$.\footnote{This is exactly the same as the stability analysis of ODEs.} In the long-time limit as $n\to\infty$, $X_n$ has a stationary variance of
\begin{equation}
    \sigma_h^2=\frac{2h}{1-(1-h)^2}=\frac{1}{1-h/2}.\notag
\end{equation}
Hence, ULA produces greater variance, thereby over-dispersing the target distribution, with a variance bias of $h/2+\mathcal{O}(h^2)$. Its stationary lag-one correlation is $1-h$. Decreasing $h$ reduces bias relative to the target distribution; however, it increases correlation per iteration. 

For a Gaussian target $\mathcal{N}(0,\Sigma)$ and $C=I$, stability is controlled by the largest eigenvalue of $\Sigma^{-1}$. A single step size must resolve the narrowest direction (smallest variance), while movement in the widest direction (largest variance) is slow. Choosing $C\approx \Sigma$ whitens all directions, allowing $h$ to be larger, and is the simplest demonstration of preconditioning seen in Section \ref{sec:effective_samplers}.

As an alternative, one can also consider the Leimkuhler--Matthews \cite{Leimkuhler2012} discretization as follows
\begin{algorithmidea}{Leimkuhler--Matthews Unadjusted Langevin Algorithm (LM-ULA)}
    Let $X_n$ be the current state of the chain, and $\xi_{n-1}$ be the noise of the previous iteration (with initial value $\xi_0\sim\mathcal{N}(0,I)$).
    \begin{enumerate}
        \item Propose the new point according to the Leimkuhler--Matthews discretization
        \begin{equation}
            X_{n+1}=X_n-hM\nabla U(X_n)+\frac{\sqrt{2hM}}{2}\left(\xi_{n} + \xi_{n-1}\right),\quad \xi_{n}\sim\mathcal{N}(0, I).\notag
        \end{equation}
        \item Set the new state as $X_{n+1}$.
    \end{enumerate}
\end{algorithmidea}
The Leimkuhler--Matthews discretization offers advantages over the Euler--Maruyama discretization by providing better accuracy for the true SDE, with the minor added cost of storing the noise from the previous iteration. However, most work on unadjusted or adjusted methods typically focuses on the Euler--Maruyama discretization, as it makes analysis and the addition of a Metropolis adjustment easier. 

\begin{figure}
    \centering
    \includegraphics[width=0.9\linewidth]{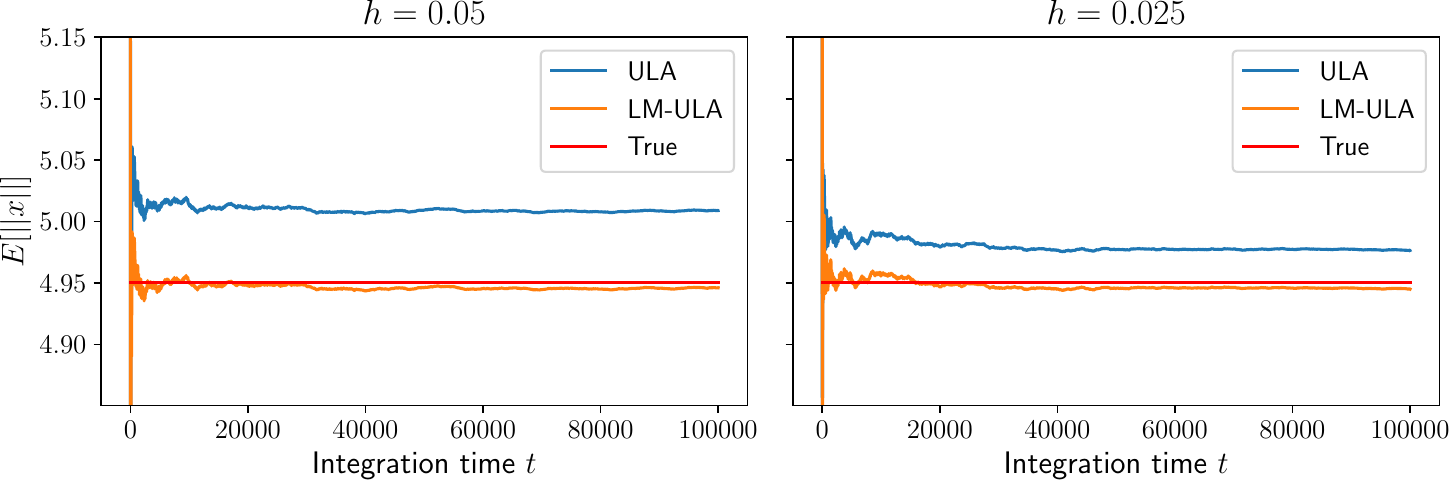}
    \caption{Discretization bias for different step sizes of ULA and LM-ULA to compute $\mathbb{E}[\|x\|]$ for $x\sim\mathcal{N}(0,I)$ and $x\in\mathbb{R}^{25}$.}\vspace{-1em}
    \label{fig:lm_ula}
\end{figure}

The result of using Euler--Maruyama versus Leimkuhler--Matthews discretization can be seen in Figure~\ref{fig:lm_ula}. Importantly, we see that as $h$ decreases, the sampler's bias decreases, which is crucial when using unadjusted sampling methods. We also see the benefit of using an integrator with higher-order accuracy; indeed, the bias from Leimkuhler--Matthews is far better than that of ULA for the same step size.

\subsection{Unadjusted HMC}
For sampling, the standard numerical integration scheme one uses is the leapfrog method \cite{Verlet1967, Strang1968}
\begin{align}
\begin{split}
    p_{n+1/2}&=p_n-\frac{h}{2}\nabla U(x_n),\\
    x_{n+1}&=x_n + hM^{-1}p_{n+1/2},\\
    p_{n+1}&=p_{n+1/2}-\frac{h}{2}\nabla U(x_{n+1}).\notag
\end{split}
\end{align}
For simplicity, and as will be made clear in the next section, define
\begin{align}
    A(x,p)= \bigg(x + hM^{-1}p,\ p\bigg)\qquad B(x,p)=\left(x,\ p-\frac{h}{2}\nabla U(x)\right).\notag
\end{align}
Then one leapfrog step consists of the composition $B\circ A\circ B$. Importantly, a leapfrog step does not give the true path of the Hamiltonian trajectories. As such, if leapfrog trajectories are used without Metropolis correction, the chain targets a perturbed distribution whose discrepancy depends on $h$ under suitable stability and ergodicity conditions.

\begin{figure}
    \centering
    \includegraphics[width=0.9\linewidth]{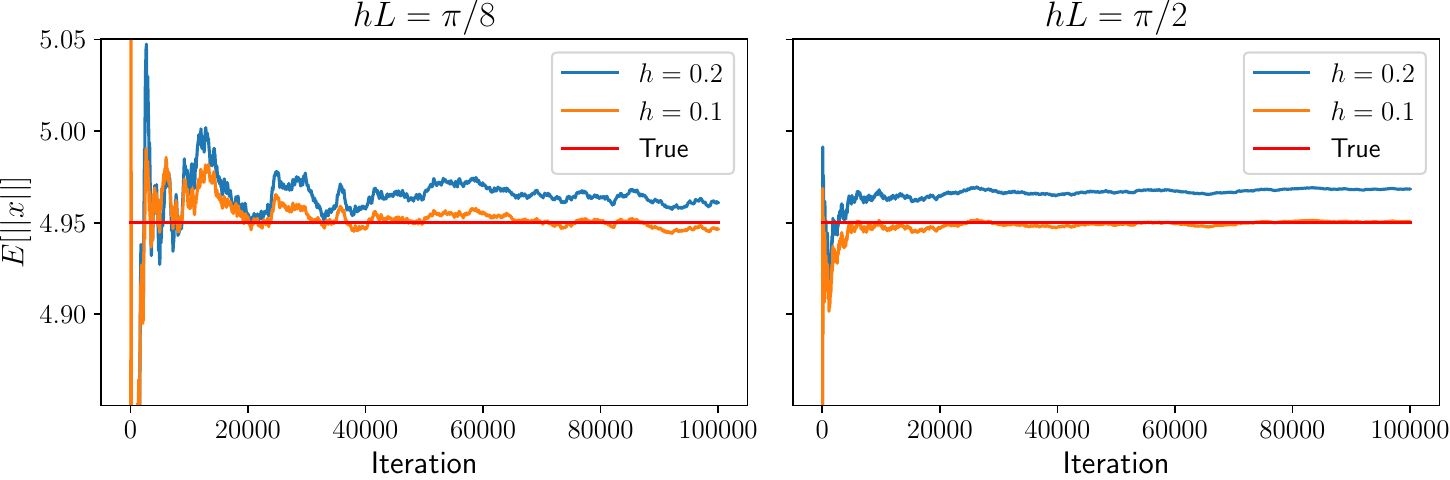}
    \caption{Discretization bias for different $h$ for unadjusted HMC using the leapfrog iteration to compute $\mathbb{E}[\|x\|]$ for $x\sim\mathcal{N}(0,I)$ and $x\in\mathbb{R}^{25}$.}\vspace{-1em}
    \label{fig:uhmc_gaussian}
\end{figure}

\begin{algorithmidea}{Unadjusted Hamiltonian Monte Carlo (UHMC)}
    Let $x_i$ be the current state of the Markov chain.
    \begin{enumerate}
        \item Draw a fresh momentum $p_i\sim\mathcal{N}(0,M)$
        \item Evolve using $L$ leapfrog steps $(B\circ A\circ B)^L(x_i,p_i)=(x_{i+1},p_{i+1})$
        \item Set the new state as $x_{i+1}$.
    \end{enumerate}
\end{algorithmidea}
Similar to exact HMC, we always accept the new state $x_{i+1}$. Figure~\ref{fig:uhmc_gaussian} shows the discretization bias of unadjusted HMC.

\begin{remark}[Choosing numerical integrators]
The leapfrog integrator is the most common, and perhaps the default, integrator for HMC. Its wide use is due to its volume-preserving nature (useful for sampling) and its reversibility under momentum negation (important for Metropolis corrections). However, many numerical integrators for Hamiltonian dynamics exist that improve the local integration accuracy \cite{BouRabee2018, Leimkuhler2005} or preserve energy \cite{Bilbao2023}. 

In practice, we find that volume-preserving integrators outperform energy-preserving integrators for sampling (its usefulness for optimization is another question). Although higher-order symmetric splitting integrators can also be volume-preserving and reversible, leapfrog remains the standard choice because it provides a particularly favorable balance between accuracy, stability, and computational cost.
\end{remark}

\subsection{Underdamped Langevin dynamics}
Underdamped Langevin dynamics naturally split into exactly solvable subflows:
\begin{align}
\begin{split}
    \mathcal{A}:&\quad \dot{x}=M^{-1}p,\quad \dot{p}=0,\\
    \mathcal{B}:&\quad \dot{x}=0,\quad \dot{p}=-\nabla U(x),\\
    \mathcal{O}:&\quad dp = -\gamma pdt + \sqrt{2\gamma M}dW_t.\notag
\end{split}
\end{align}
With corresponding discretization maps \cite{Bussi2007, SKEEL2002, Melchionna2007}
\begin{align}
\begin{split}
    A(x,p)&= \bigg(x + hM^{-1}p,\ p\bigg)\\ 
    B(x,p)&=\left(x,\ p-\frac{h}{2}\nabla U(x)\right)\\
    O(x,p)&=\bigg(x,e^{-\gamma h/2}p+\sqrt{1-e^{-\gamma h}}M^{1/2}\xi\bigg),\quad \xi\sim\mathcal{N}(0,I).
\end{split}
\end{align}
Classic discretizations of underdamped Langevin dynamics consider palindromic compositions such as BAOAB or OBABO, yielding weak second-order approximations while providing an exact treatment of momentum refreshment. In practice, the best splitting depends on the application: in molecular sampling, BAOAB has configurational advantages \cite{Leimkuhler2012}, while OBABO supports a reversible Metropolis proposal.

\begin{figure}
    \centering
    \includegraphics[width=0.9\linewidth]{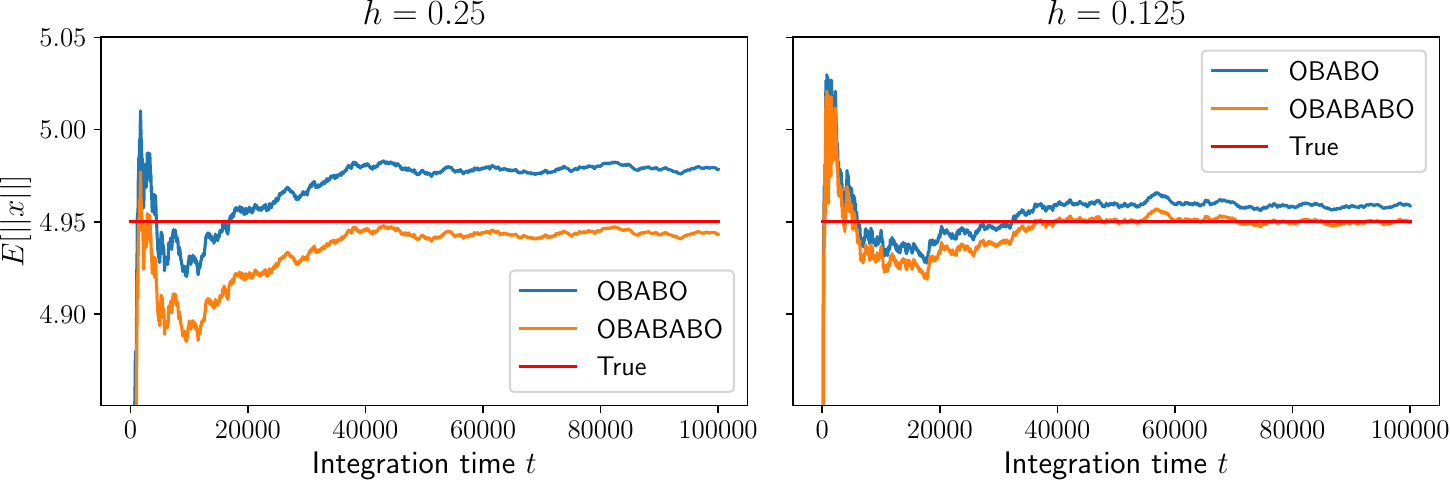}
    \caption{Discretization bias for different step sizes of OBABO and OBABABO to compute $\mathbb{E}[\|x\|]$ for $x\sim\mathcal{N}(0,I)$ and $x\in\mathbb{R}^{25}$.}\vspace{-1em}
    \label{fig:obabo_obababo}
\end{figure}

\begin{algorithmidea}{OBABO Underdamped Langevin (OBABO)}
    Let $(x_n,p_n)$ be the position and momentum of the chain, and $h$ be the step size.
    \begin{enumerate}
        \item Perform the partial momentum refreshment $(\widetilde x_n,\widetilde p_n)=O(x_n,p_n)$, that is
        \begin{equation}
            \widetilde x_n = x_n,\quad  \widetilde p_n = \eta p_n+\sqrt{1-\eta^2}M^{1/2}\xi_1,\qquad \xi_1\sim\mathcal{N}(0, I),\notag
        \end{equation}
        where $\eta = e^{-\gamma h/2}$.
        \item Evolve according to the Hamiltonian flow BAB: $(x_n',p_n')=BAB(\widetilde x_n,\widetilde p_n)$.
        \item Perform the last partial momentum refreshment $(x_{n+1},p_{n+1})=O(x_n',p_n')$
        \begin{equation}
            x_{n+1} = x_n',\quad  p_{n+1} = \eta p_n'+\sqrt{1-\eta^2}M^{1/2}\xi_2,\qquad \xi_2\sim\mathcal{N}(0, I).\notag
        \end{equation}
        \item Set the new state as $(x_{n+1}, p_{n+1})$.
    \end{enumerate}
\end{algorithmidea}
It is worth noting that the two O steps are exact, whereas the BAB step introduces the discretization error. As such, Metropolis adjustments for underdamped Langevin mechanics are used to "fix" this error. Figure~\ref{fig:obabo_obababo} shows the discretization bias introduced by OBABO and OBABABO; the latter provides a slightly more accurate integrator. Pseudocode for the OBABO and OBABABO discretization can be found in Appendix~\ref{app:obabo_obababo}.

\subsection{When unadjusted methods are sensible}
By nature, unadjusted methods introduce a bias in the invariant distribution. That choice can be rational when the Metropolis correction is more costly than the bias is harmful. Examples include:
\begin{itemize}
    \item Approximate posterior prediction in very large models (e.g., Bayesian neural networks)
    \item Stochastic gradients for which the exact proposal density is unavailable or noisy
    \item Early exploratory computations where the dominant uncertainty is model misspecification rather than numerical bias
    \item Monte Carlo optimization of marginal likelihoods or other expectations that cannot be evaluated analytically. In such settings, an unadjusted sampler can provide inexpensive approximate draws to estimate the required integral at each optimization step, provided that the resulting discretization bias is small relative to the statistical and optimization errors.
\end{itemize}
The key requirement is not that the bias be zero, but that it can be controlled relative to the objective of interest. When using an unadjusted method, one must therefore consider how to reduce the bias introduced by discretization. Several approaches are available:
\begin{itemize}
    \item Use a smaller step size -- This should be done in any case to ensure that the numerical scheme remains stable. Results should also be compared across several sufficiently small step sizes to verify that the conclusions are not sensitive to the particular choice of $h$.
    \item Use a preconditioning matrix -- As further explored in Section~\ref{sec:effective_samplers}, taking $M$ as the covariance matrix of $\pi$ can dramatically increase the step size while preserving numerical stability. However, when the problem's dimensionality is high ($d>1000$), this may not be computationally feasible.
    \item Use a better discretization -- As shown in Figure~\ref{fig:lm_ula} and Figure~\ref{fig:obabo_obababo}, a more accurate numerical scheme can reduce the bias at a fixed step size $h$. The increasing availability of reliable software and code-generation tools (LLMs) has also made higher-order schemes easier to implement in practice.
\end{itemize}

\section{Metropolis--Hastings Adjustments}
Let a proposal kernel have density $q(z,z')$ on a state space that may include auxiliary variables. Metropolis--Hastings \cite{Metropolis1953, Hastings1970} accepts a proposal $z'$ from $z$ with probability
\begin{equation}\label{eq:general_acceptance_probability}
    \alpha(z,z')=\min\left(1,\frac{\widetilde{\pi}(z')q(z',z)}{\widetilde{\pi}(z)q(z,z')}\right).
\end{equation}
Adding this accept--reject stage removes the stationary discretization bias but introduces possible rejections and a resulting loss of sampling efficiency. However, this introduces a new mechanism for inefficiencies. If $h$ is too large, proposals are rejected (as the discretization is far from the true dynamics), and the chain rarely moves. If $h$ is too small, then more proposals are accepted, but more iterations are required to explore the state space effectively. 

\subsection{Hamiltonian Monte Carlo}
When one refers to HMC, one typically refers to the adjusted version as follows.
\begin{algorithmidea}{Hamiltonian Monte Carlo (HMC)}
    Let $x_i$ be the current state of the Markov chain
    \begin{enumerate}
    \item Draw a fresh momentum $p_i\sim\mathcal{N}(0,M)$
    \item Evolve using $L$ leapfrog steps $(B\circ A\circ B)^L(x_i,p_i)=(x_{i+1},p_{i+1})$
    \item Accept $x_{i+1}$ with probability 
    \begin{equation}
        \alpha(z_{i+1},z_i)=\min\left(1,{\exp\big[-H(x_{i+1}, p_{i+1}) +H(x_i,p_i)\big]}\right).\notag
    \end{equation}
    Upon rejection, retain $x_{i}$.
    \item If one wants to keep the momentum, accept $(x_{i+1}, -p_{i+1})$ with probability $\alpha(z_{i+1},z_i)$, and return $(x_i,p_i)$ otherwise.
\end{enumerate}
\end{algorithmidea}

The step size $h$ and number of steps $L$ play different roles. The step size controls integration accuracy (i.e., $H(z_{i+1})-H(z_i)$) and thus acceptance rate, while the integration time $T=Lh$ controls the distance traveled. As seen in the exact HMC case, if $T$ is too long, the trajectory can pass the quarter-period point of maximal displacement and begin returning toward its starting position; at $T=2\pi$, it returns exactly to the initial position. Such overly long trajectories waste computation on additional gradient evaluations. 

\subsection{The No-U-Turn Sampler}
The No-U-Turn Sampler (NUTS) \cite{hoffman2014no} addresses the problem of oscillating and doubling-back trajectories by dynamically building a Hamiltonian trajectory until continued expansion would begin to double back. This is performed by using a two-sided no-u-turn condition, which in the identity Gaussian example (Section~\ref{sec:gaussian_lab}) occurs at the exception quarter-period HMC sampler.\footnote{While this gives useful intuition for the stopping rule, the realized NUTS stopping time and selected proposal are random and need not equal the quarter-period state.}

The main advantage of NUTS is that it avoids choosing a fixed $L$ for each target. This is, however, at the cost of branching instructions and a variable number of gradient evaluations. A binary tree of depth $D$ can contain up to order $2^D$ leapfrog steps: the computational cost is linear in the number of steps actually constructed, while the maximum permitted work grows exponentially with the maximum tree depth. For the interested reader, we explain the No-U-Turn condition and how NUTS uses it to construct efficient proposals below.

\textbf{One-sided No-U-Turn Condition.} Let $X_0$ be the initial condition, and $(X_t, P_t)$ the result of (numerical or exact) integration along the Hamiltonian flow. The basic No-U-Turn condition is
\begin{equation}
    (X_t-X_0)\cdot P_t<0.\notag
\end{equation}
Going back to the exact HMC, we see that
\begin{equation}
    \mathbb{E}[(X_t-X_0)\cdot P_t] = \sin t(1-\cos t)\mathbb{E}\|X_0\|^2+\sin t\cos t\mathbb{E}\|P_0\|^2+(\cos 2t-\cos t)\mathbb{E}[X_0\cdot P_0].\notag
\end{equation}
At stationarity, $X_0,P_0\sim\mathcal{N}(0,I)$, so
\begin{equation}
    \mathbb{E}[(X_t-X_0)\cdot P_t]=d\sin t.\notag
\end{equation}
Therefore, the expected one-sided No-U-Turn criterion remains positive for $t\in(0,\pi)$, and first reaches $0$ at $t=\pi$, where the samples oscillate (seen in Figure~\ref{fig:exact_hmc}). However, we would like to identify the ideal case where $t=\pi/2$. 

\textbf{Two-sided No-U-Turn Condition.} Currently, we only consider integrating the Hamiltonian dynamics forwards in time. However, we can also consider integrating the Hamiltonian dynamics backward in time (which is equivalent to setting $h\to -h$ in the leapfrog steps). Let $X_{+t}$ and $X_{-t}$ denote the forward and backward trajectories for time $t$, respectively. Explicitly, we have 
\begin{align}
    X_{\pm t} &= X_0\cos t \pm P_0\sin t,\notag\\
    P_{\pm t} &= P_0\cos t \mp X_0\sin t.\notag
\end{align}
The two-sided condition is then
\begin{equation}
    (X_{+t}-X_{-t})\cdot P_{-t}\geq 0,\quad\text{and}\quad (X_{+t}-X_{-t})\cdot P_{+t}\geq 0.\notag
\end{equation}
In particular, this asks when the forward and backward trajectories begin the U-turn. Performing the same computations as above, we see that
\begin{equation}
    \mathbb{E}[(X_{+t}-X_{-t})\cdot P_{-t}]=\mathbb{E}[(X_{+t}-X_{-t})\cdot P_{+t}]=d\sin(2t).\notag
\end{equation}
As such, the first positive zero of the expected symmetric no-u-turn
criterion occurs at $t=\pi/2$, coinciding with the quarter-period flow
that produces an independent exact-HMC position proposal for the
identity Gaussian.

\subsection{Metropolis-Adjusted Langevin Algorithm}
The unadjusted Langevin proposals \eqref{eq:ula} take a Gaussian form
\begin{equation}
    q(x,\cdot)=\mathcal{N}(x-hM\nabla U(x), 2hM).
\end{equation}
The Metropolis-Adjusted Langevin Algorithm (MALA) then applies the Metropolis correction with acceptance probability \eqref{eq:general_acceptance_probability}. The algorithm is as follows
\begin{algorithmidea}{Metropolis-Adjusted Langevin Algorithm (MALA)}
Let $X_n$ be the current state of the chain
\begin{enumerate}
    \item Draw $X_{n+1}\sim \mathcal{N}(X_n-hM\nabla U(X_n), 2hM)$, equivalently,
    \begin{equation}
        X_{n+1}=X_n - hM\nabla U(X_n)+\sqrt{2hM}\xi_n,\qquad \xi_n\sim\mathcal{N}(0,I).\notag
    \end{equation}
    \item Accept $X_{n+1}$ with probability
    \begin{equation}
        \alpha(X_{n+1}, X_n)=\min\left(1,\frac{\exp(-U(X_{n+1}))q(X_{n+1},X_{n})}{\exp(-U(X_{n}))q(X_{n},X_{n+1})}\right),\notag
    \end{equation}
    otherwise, return $X_n$.
\end{enumerate}
\end{algorithmidea}

MALA's strengths are simplicity, low memory, predictable cost, and easy vectorization across chains. However, these strengths are not unique to MALA and are shared by Metropolized versions of underdamped Langevin dynamics, with the added cost of doubling the space requirement. MALA can be highly effective for targets that are approximately Gaussian after a fixed preconditioning step. Its weaknesses arise when the local Gaussian approximation changes sharply with position.

\subsection{Metropolis-Adjusted Underdamped Langevin}
There are many variants of adjusted Underdamped Langevin mechanics, e.g., GHMC \cite{Horowitz1991, Kennedy1996}, MALT \cite{durand2023, Ruiz2023}, and MAKLA \cite{BouRabee2024, chok2026}. In this section, we focus on MAKLA, which is as follows:
\begin{algorithmidea}{One OBABO step MAKLA}
    Let $(x_n,p_n)$ be the current position and momentum of the chain, and let $h$ be the step size.
    \begin{enumerate}
    \item Perform the first partial momentum refreshment $(\widetilde x_n,\widetilde p_n)=O(x_n,p_n)$, that is,
    \begin{equation}
        \widetilde x_n = x_n,\quad  \widetilde p_n = \eta p_n+\sqrt{1-\eta^2}M^{1/2}\xi,\qquad \xi\sim\mathcal{N}(0, I),\notag
    \end{equation}
    where $\eta = e^{-\gamma h/2}$.
    \item Evolve according to the Hamiltonian flow BAB:
    \begin{equation}
        (x_n',p_n')=BAB(\widetilde x_n,\widetilde p_n).\notag
    \end{equation}
    \item Compute the Hamiltonian error accumulated during the BAB step,
    \begin{equation}
        \Delta=H(x_n',p_n')-H(\widetilde x_n,\widetilde p_n).\notag
    \end{equation}
    \item With probability $\min(1,\exp(-\Delta))$, set
    \begin{equation}
        (\overline x_n,\overline p_n)=(x_n',p_n');\notag
    \end{equation}
    otherwise, set
    \begin{equation}
        (\overline x_n,\overline p_n)=(\widetilde x_n,-\widetilde p_n).\notag
    \end{equation}
    \item Perform the final partial momentum refreshment
    \begin{equation}
        (x_{n+1},p_{n+1})=O(\overline x_n,\overline p_n).\notag
    \end{equation}
\end{enumerate}
\end{algorithmidea}
MAKLA occupies a middle ground between MALA and NUTS. It uses regular proposals suitable for GPUs, but momentum persistence can produce longer, more coherent movement than MALA. One weakness is that a poorly chosen global step size or refreshment rate can still lead to slow exploration or rejection-induced reversals. Moreover, if the proposal is rejected, the negation of the momentum means that the sampler backtracks, reducing sampling efficiency.\footnote{MALT avoids this backtracking by fully refreshing the momentum between successive trajectories; however, the (O)-substeps within the OBABO integrator still perform partial momentum refreshment. MAKLA instead preserves momentum across iterations and must reverse it following a rejection to maintain the required reversibility of the Markov transition.}

In practice, we find that using $L=1$ and keeping $h$ small, so that the acceptance rate is $\geq 90\%$, often works well. This ensures that the Metropolized sampler stays close to the true continuous dynamics. Similar to MALA and NUTS, MAKLA (and variants of adjusted Underdamped Langevin) work well for Gaussian-like distributions after a fixed preconditioning. However, they can often fail for funnel-like distributions.

\section{From valid kernels to effective samplers}\label{sec:effective_samplers}
The above Metropolis-adjusted (or unadjusted) kernels can preserve $\pi$ (or $\pi_h$) exactly but have long mixing times. The usual reason is typically due to two separate failure modes:
\begin{enumerate}
    \item Global anisotropy: Distributions which are Gaussian-like often come with directions with large variance and other directions with small variance. The direction with small variance requires smaller step sizes to keep the numerical discretization error small. In this case, the Hessian of the distribution is approximately the same across the state space; however, it has a large dynamic range of eigenvalues.
    \item Local multiscale geometry: Distributions may contain regions of high local curvature and other regions of low local curvature. The high-curvature regions force small step sizes, at the cost of poor exploration in the low-curvature regions. In this case, the eigenvalues of the Hessian change dramatically across the state space. 
\end{enumerate}
These two issues motivate two classes of preconditioning strategies. In the first case, a global preconditioning method proves to be extremely useful. In the second case, one must resort to local preconditioning methods, such as using Riemannian metrics or time-rescaled dynamics.  

\subsection{Interlude -- Multimodal distributions}
Parallel tempering schemes \cite{Swendsen1986,Hukushima1996,Earl2005} provide a useful approach for sampling from multimodal distributions. The principal difficulty in such problems is that a Markov chain may remain trapped near one mode for a long time before crossing the low-probability region separating it from another mode.

Parallel tempering runs several Markov chains simultaneously, with the $k$-th chain targeting the tempered density
\begin{equation}
    \pi_{\beta_k}(x) \propto \exp\left(-\beta_k U(x)\right), \qquad 1=\beta_0>\beta_1>\cdots>\beta_K>0,\notag
\end{equation}
where $\beta_k$ is the inverse temperature. For example, an overdamped
Langevin diffusion with invariant density $\pi_{\beta_k}$ is
\begin{equation}
    dX_t^k = -\nabla U(X_t^k)dt + \sqrt{\frac{2}{\beta_k}}dW_t^k.\notag
\end{equation}
Equivalently, after a deterministic rescaling of time, one may write
\begin{equation}
    dX_t^k = -\beta_k\nabla U(X_t^k)\,dt + \sqrt{2}\,dW_t^k.\notag
\end{equation}

The cold chain, corresponding to $\beta_0=1$, targets the distribution of interest. For smaller values of $\beta_k$ (the hot chains), the potential landscape flattens, and noise dominates the dynamics, reducing the effective energy barriers between modes. The hotter chains can therefore move more readily between modes and through regions that have low probability under the original target.

Periodic Metropolis exchanges between neighboring temperatures allow states discovered by the hotter chains to propagate back to the cold chain. In particular, a proposed exchange of states $x_i$ and $x_j$ between chains with inverse temperatures $\beta_i$ and $\beta_j$ is accepted with probability
\begin{equation}
    \min\left\{1, \exp\left[(\beta_i-\beta_j)\bigl(U(x_i)-U(x_j)\bigr)\right]\right\}.\notag
\end{equation}
The performance of parallel tempering depends strongly on the choice of
temperature ladder and the frequency of exchange attempts, and adaptive
schemes have been developed to tune these quantities automatically
\cite{Miasojedow2013,Vousden2015,Syed2021}. Although naturally
parallelizable, the method requires simulating several replicas and can
therefore be computationally expensive.

In practice, we find that the global preconditioning strategy described below can substantially reduce mode trapping, particularly when several chains are run in parallel. By constructing a common preconditioner from information pooled across many parallel chains (adaptive samplers), the sampler can capture variation both within and between the regions explored by different chains, thereby improving global exploration of the state space. In a multimodal setting, an adaptively estimated covariance may therefore encode information about the relative locations and scales of multiple modes, increasing the frequency of transitions between them. This approach does not eliminate the intrinsic difficulty of crossing low-probability barriers, but it can considerably reduce the tendency of individual chains to remain confined to a single mode.

\subsection{Global preconditioning and whitening}
In all three samplers described in the previous section (which also apply to NUTS), there were matrices $M$ for HMC and underdamped Langevin dynamics, and $C$ for overdamped Langevin dynamics. Previously, we treated $M$ and $C$ as the identity matrix, but this is exactly where global preconditioning comes into play. 

The simplest strategy for preconditioning is to set the Hessian at the mode.
\begin{algorithmidea}{Hessian at MAP Preconditioning}
\begin{enumerate}
    \item Use an optimization routine, like L-BFGS-B, to find the mode (equivalently, the MAP) $\widehat{x}$
    \item Use an autodifferentiation package to compute the Hessian of the negative log-density at the mode
    \begin{equation}
        \widehat{H}=\nabla^2 U(\widehat{x})=-\nabla^2\log \pi(\widehat{x}).\notag
    \end{equation}
    \item Set $M=\widehat{H}$ for HMC and underdamped Langevin, and $C=\widehat{H}^{-1}$ for overdamped Langevin.
\end{enumerate}    
\end{algorithmidea}

This simple strategy, which adds only an extra computational cost during sampler initialization, is powerful when the posterior is locally Gaussian and the mode is representative. However, this strategy can be misleading when a Gaussian approximation at the mode is not representative. \texttt{R} code to perform the Hessian at MAP preconditioning for NUTS can be found in Appendix~\ref{app:RCode}. 

Another strategy dynamically computes this preconditioning matrix $M$ (equivalently, $C=M^{-1}$ for overdamped Langevin). Namely, during the \emph{burn-in} phase, one can sequentially update a covariance estimate $C_K$ from the historical states of the Markov chain and set $M_K$ to its regularized inverse. Suppose we are at iteration $K\geq 2$ of the burn-in phase. We update the running mean and covariance using the multivariate Welford-type recursion \cite{Welford1962, Chan1983}
\begin{align}
    \begin{split}\label{eq:adaptive_estimations}
    \mu_K &= \frac{1}{K}X_K + \left(1-\frac{1}{K}\right)\mu_{K-1},\\
    C_K&= \frac{1}{K}(X_K-\mu_{K-1})\otimes (X_K-\mu_{K}) + \left(1-\frac{1}{K}\right) C_{K-1},\\
    M_K&=(C_K+\varepsilon I)^{-1}.
    \end{split}
\end{align}
We initialize $\mu_1=\widehat{x}$, $M_1=\widehat{H}$, and $C_1=M_1^{-1}=\widehat{H}^{-1}$, where $\varepsilon>0$ is a small regularization parameter. Because $C_1$ is nonzero, $C_K$ is a MAP-initialized running covariance estimate rather than the ordinary empirical covariance, and the contribution of the initialization decays proportionally to $1/K$. Thus, $\mu_K$ approximates the mean of the distribution, while $M_K$ approximates its inverse covariance. 

\begin{algorithmidea}{Finite Adaptive MCMC}
\begin{itemize}
    \item For HMC and underdamped Langevin, the $K$-proposal during burn-in is constructed using $M_K$ in place of $M$, or $C_K$ in place of $C$ for overdamped Langevin dynamics. 
    \item At the end of burn-in, the adaptation is stopped and the final estimate, denoted by $M_{\mathrm{final}}$ or $C_{\mathrm{final}}$, is held fixed throughout the sampling phase.
    \item (Optional) Include an additional stabilization period after the final adaptation before retaining samples.
\end{itemize}
\end{algorithmidea}
Using $M_K$ (or $C_K$) in place of $M$ (or $C$) allows the sampler to learn the posterior geometry dynamically, rather than relying solely on a local approximation obtained at the mode. Moreover, freezing the preconditioner ensures that the sampling phase is governed by a fixed Markov transition kernel, thus preserving $\pi$ as the invariant distribution for the Metropolis-adjusted and exact versions.

More generally, one can consider a sampler in which $C_K$ and $\mu_K$ continue to be adapted during the sampling phase, called Adaptive MCMCs \cite{Andrieu2008, Haario2001, roberts2007}. However, this requires careful analysis of the target distribution to ensure proper convergence of the Markov chain. Moreover, adapting a dense covariance matrix $C_K$ can incur additional overhead, as a Cholesky decomposition must be performed at every iteration. In practice, we find that learning $C_K$ during the burn-in phase and fixing it during the sampling phase often provides the convergence benefits of adaptive MCMC without additional computational overhead.

In cases where the dimension is too large, so a full Hessian is computationally impractical, a diagonal approximation or block-diagonal approximation can also improve performance.

\begin{figure}
    \centering
    \includegraphics[width=0.95\linewidth]{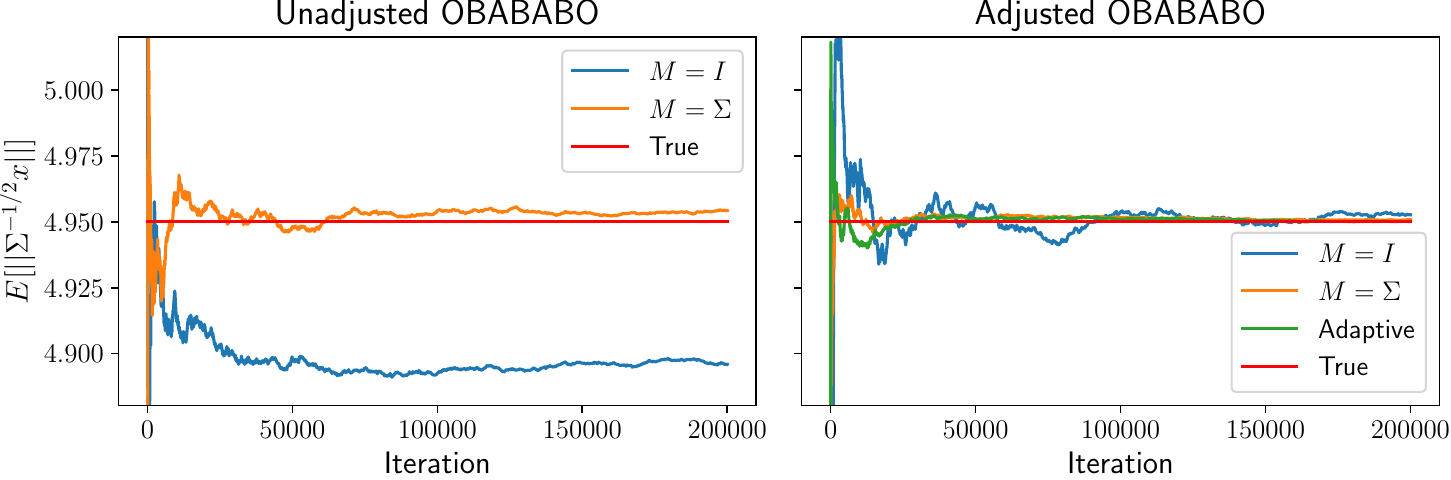}
    \caption{Evolution of the empirical estimate of $\mathbb{E}[\|\Sigma^{-1/2}x\|]$, where $x\sim \mathcal{N}(0,\Sigma)$, $\lambda_{\min}(\Sigma)=10^{-2}$ and $\lambda_{\max}(\Sigma)=10^{2}$, and $x\in\mathbb{R}^{25}$. The left panel compares unadjusted OBABABO with no preconditioning, $M=I$, and inverse-Hessian preconditioning, $M=\Sigma$, using $h=0.25$. The right panel compares the corresponding Metropolis-adjusted samplers with an adaptively estimated preconditioner initialized at $M_1=I$; in each case, the step size is tuned to achieve an acceptance rate of $0.9$. The horizontal red line denotes the exact value.}
    \label{fig:precond_gaussian}
\end{figure}

\begin{figure}
    \centering
    \includegraphics[width=0.95\linewidth]{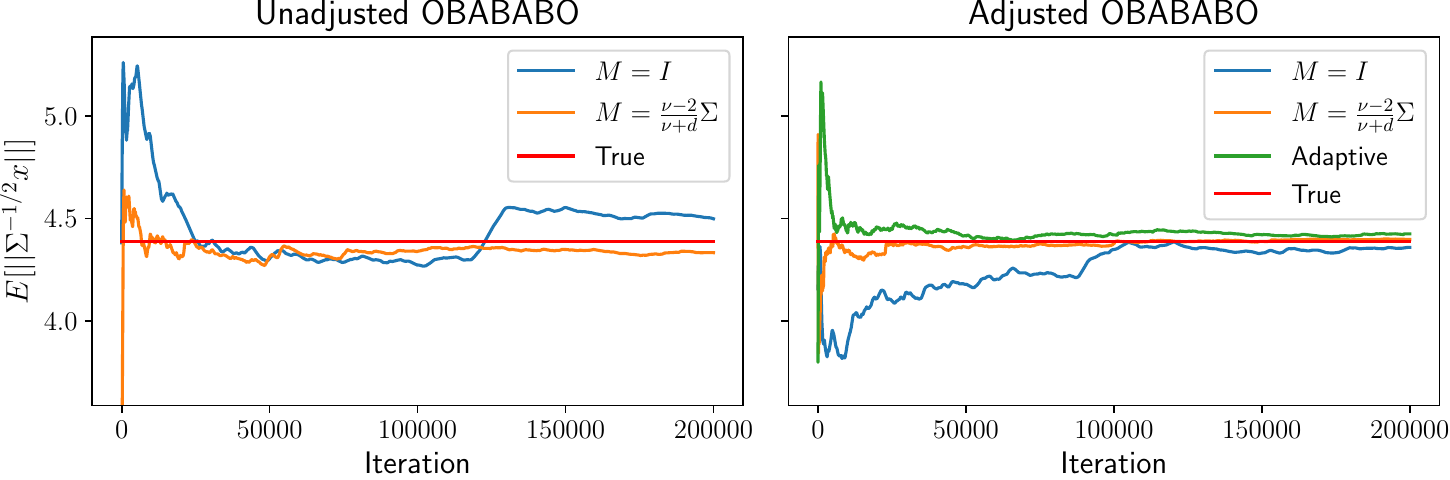}
    \caption{Evolution of the empirical estimate of $\mathbb{E}[\|\Sigma^{-1/2}x\|]$, where $x\sim t_4(0,\Sigma)$, $\lambda_{\min}(\Sigma)=10^{-2}$ and $\lambda_{\max}(\Sigma)=10^{2}$, and $x\in\mathbb{R}^{25}$. The left panel compares unadjusted OBABABO with no preconditioning, $M=I$, using $h=0.1$, and inverse-Hessian preconditioning, $M=(\nu-2)(\nu+d)^{-1}\Sigma$, using $h=0.25$. The right panel compares the corresponding Metropolis-adjusted samplers with an adaptively estimated preconditioner initialized at $M_1=I$; in each case, the step size is tuned to achieve an acceptance rate of $0.9$. The horizontal red line denotes the exact value.}\vspace{-1em}
    \label{fig:precond_student_t}
\end{figure}

The effect of preconditioning by the inverse negative Hessian at the MAP is illustrated in Figure~\ref{fig:precond_gaussian} for the Gaussian target $\mathcal{N}(0,\Sigma)$ and in Figure~\ref{fig:precond_student_t} for the Student-$t$ target $t_4(0,\Sigma)$, where $\Sigma\in\mathbb{R}^{25\times 25}$ is a dense covariance matrix with eigenvalues linearly spaced between $10^{-2}$ and $10^2$. The unadjusted methods use a step size of $h=0.25$, except for the unpreconditioned Student-$t$ sampler, corresponding to $M=I$, for which we use $h=0.1$.\footnote{With $h=0.25$, the discretization bias was large, approximately one order of magnitude greater than the true value.} For the adaptively preconditioned, Metropolis-adjusted OBABABO method, the preconditioner is initialized at the identity, $M_1=I$, and the adapted preconditioner is held fixed during the sampling phase.
\begin{itemize}
    \item For the unadjusted methods, preconditioning with the inverse Hessian at the MAP substantially reduces the bias relative to the unpreconditioned sampler. This is expected because the preconditioner reduces the effective anisotropy and the Lipschitz constant of the transformed gradient, thereby improving the numerical stability of the integrator.
    \item For the Metropolis-adjusted methods, preconditioning with the inverse Hessian at the MAP performs best, substantially outperforming both adaptive preconditioning and no preconditioning. Adaptive preconditioning performs nearly as well, while the unpreconditioned sampler converges the slowest.\\
\end{itemize}

\begin{remark}
    When the mode and its local curvature are representative of the typical set, MAP--Hessian initialization can improve the early stages of warmup. However, it may be unhelpful when the mode lies in an atypical region, the posterior is multimodal, or the local Hessian is poorly conditioned. In such cases, adapting the preconditioner using empirical covariance estimates from the evolving chains can provide a more representative approximation to the posterior geometry and thus enhance sampling performance.
\end{remark}

\begin{remark}
    Other examples of global preconditioning strategies exist \cite{clarte2022, sprungk2023} based on mean-field dynamics. These build global preconditioning matrices by running a (large) ensemble of Markov chains and use these chains to estimate some global target with respect to the target distribution, e.g., the mean or covariance. However, as noted in \cite{chok2026}, adaptive samplers perform this same task without the requirement of a large ensemble size. For more details on adaptive MCMC, we refer the reader to \cite{Liang2010, laitinen2024invitationadaptivemarkovchain}.
\end{remark}

\subsection{Neal's funnel and hierarchical geometry}
Neal's funnel \cite{Neal2003} can be written as
\begin{align}
\begin{split}\label{eq:neals_funnel}
    v &\sim \mathcal{N}(0, \sigma^2),\\
    x_i|v&\sim \mathcal{N}(0, e^{v}),\quad i=1,\ldots, d,
\end{split}
\end{align}
where $\sigma$ is a fixed constant. The log density, up to a constant, is
\begin{equation}
    \log \pi(v,x)=-\frac{v^2}{2\sigma^2}-\frac{d}{2}v-\frac{1}{2}e^{-v}\sum_{i=1}^dx_i^2.\notag
\end{equation}
As such, the $x$-block Hessian $-\nabla^2_{x_i}\log \pi (v,x)=e^{-v}I$, so its eigenvalues vary exponentially with $v$. This contrasts with the anisotropic case, in which the Hessian is nearly constant (e.g, a Gaussian distribution with a fixed covariance matrix). 

When $v$ is negative, the $x_i$ coordinates are confined to a narrow neck with large curvature, while when $v$ is positive, they occupy a wide mouth with small curvature. Thus, a global step size stable at $v=-4$ is wastefully small at $v=4$. Moreover, a global covariance matrix which averages over this variation cannot make both regions isotropic simultaneously.

The funnel is not a contrived curiosity. It is a canonical geometry of a centered hierarchical scale model. For example, the horseshoe prior
\begin{equation}
    \beta_i|\tau_0,\tau_i\sim\mathcal{N}(0,\tau_0^2\tau_i^2)\notag
\end{equation}
which creates narrow coefficient scales when the global or local shrinkage parameters are small and wide scales when they are large. For Bayesian modeling, however, given enough data, the posterior distribution may become approximately Gaussian due to the Bernstein--von Mises theorem, which may explain why a centered parameterization can be harmless in large-data regimes.

\begin{remark}
A useful approach would be to consider non-centered parameterization, in which case one writes
\begin{equation}
    \beta_i=\tau_0\tau_iz_i,\qquad z_i\sim\mathcal{N}(0,1).\notag
\end{equation}
This removes the prior funnel in $(z,\tau)$ coordinates; however, the likelihood can introduce new complications. Moreover, in more complex hierarchical models (e.g., when the horseshoe prior is applied to finite differences of the $\beta_i$ coefficients), it may not always be computationally practical. 

If one doesn't know how to remove the funnel-like geometry, \cite{Gorinova_2020} provides a demonstration on how this reparameterization can be performed automatically. 
\end{remark}

\subsection{Position dependent metrics}
So far, we have considered only the case where the preconditioning matrix $M$ is fixed. However, allowing $M$ to vary with the chain's current position can alleviate issues with funnel-like geometries. Particularly for HMC and underdamped Langevin, the cost is that building such schemes (whether unadjusted or adjusted) can be computationally impractical.

The simplest form of position-dependent preconditioning comes through the following overdamped Langevin dynamics \cite{Xifara2014, Roy2022, Girolami2011, hsieh2018mirrored}
\begin{equation}
    dX_t = -C(X_t)\nabla U(X_t)\,dt+\nabla\cdot C(X_t)\,dt+\sqrt{2C(X_t)}\,dW_t,
\end{equation}
where $(\nabla\cdot C)_i=\sum_j\partial_j C_{ij}$, and $C(x)$ may be chosen as a regularized positive-definite approximation to the inverse Hessian of the negative log-density at $x$. In this case, the continuous-time dynamics exactly target $\pi$ as the invariant distribution. The corresponding discretized dynamics is therefore
\begin{equation}
    X_{n+1}=X_n - h\Big(C(X_n)\nabla U(X_n)-\nabla\cdot C(X_n)\Big)+\sqrt{2h C(X_n)} \xi_n,\qquad \xi_n\sim\mathcal{N}(0,I).
\end{equation}

To add a Metropolis adjustment, define
\begin{align}
    \mu(x)&=C(x)\nabla U(x)-\nabla\cdot C(x),\quad\text{and}\notag\\
    \widetilde\mu(x)&=C(x)\nabla U(x).\notag
\end{align}
We can then consider the two proposal kernels
\begin{equation}
    q_h(x,\cdot)=\mathcal{N}\bigl(x-h\mu(x),2hC(x)\bigr),\qquad
    \widetilde q_h(x,\cdot)=\mathcal{N}\bigl(x-h\widetilde\mu(x),2hC(x)\bigr),\notag
\end{equation}
with the usual Metropolis adjustment computed using the proposal density that is actually used. While the first proposal converges to the displayed position-dependent Langevin SDE as $h\to0$, in practice we find that adding the extra term $\nabla \cdot C(x)$ incurs additional computational overhead with little empirical benefit. Omitting this term changes the proposal and its small-step diffusion limit, but a Metropolis correction based on $\widetilde q_h$ still preserves $\pi$ exactly.

\begin{remark}[Sampling from Constrained Distributions]
Interestingly, this state-dependent preconditioned Langevin can also be used to sample from constrained distributions \cite{ChokPetzina2026}. In this case, the $C(X_t)$ informs the sampler how far $X_t$ is from the boundary of the constrained domain. The closer it is to the boundary, the sampler slows down; the farther away, the faster.
\end{remark}

\begin{remark}
Many tamed Langevin algorithms \cite{Atchad2006, Brosse2019, Hutzenthaler2012, Sabanis2013, Johnston2025} replace the original drift by a step-size-dependent bounded or attenuated drift in order to prevent instability for superlinear gradients. The resulting fixed-step chain is generally biased, although under suitable assumptions its invariant law approaches the target as the step size decreases.
\end{remark}

\begin{figure}
    \centering
    \includegraphics[width=0.95\linewidth]{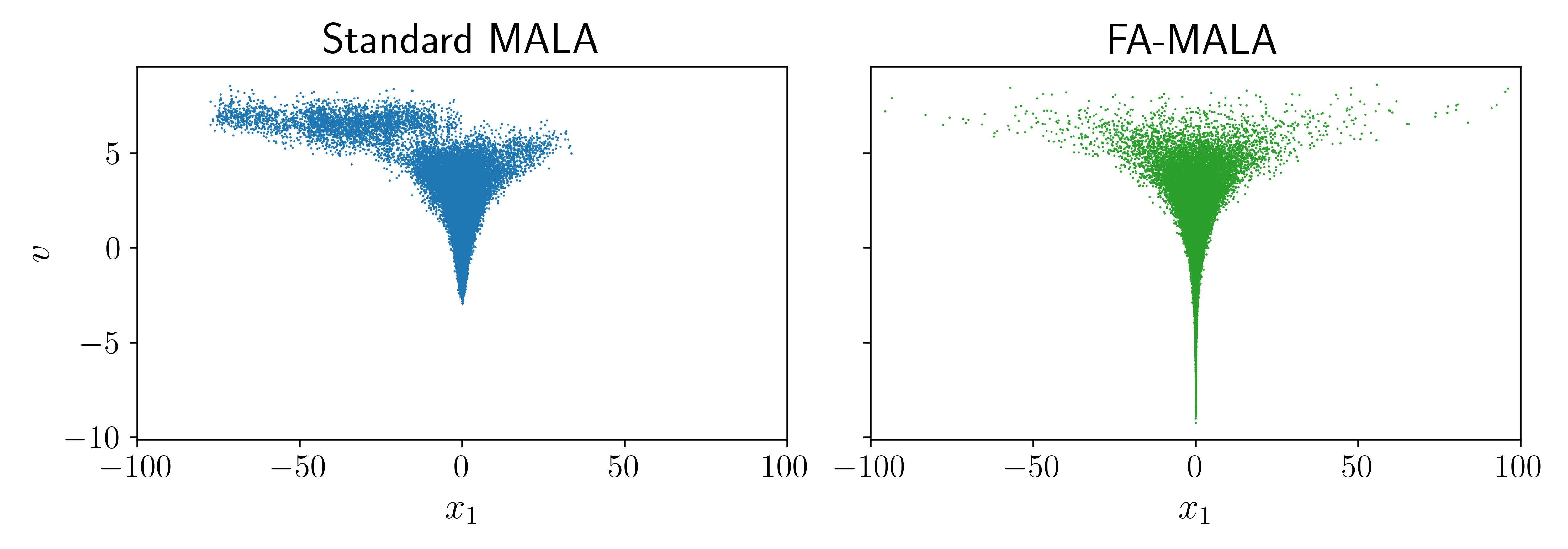}
    \caption{Samples from Neal’s funnel projected onto the $(x_1,v)$-coordinates for standard MALA (left) and FA-MALA (right).}
    \label{fig:nealsfunel}\vspace{-1em}
\end{figure}

\paragraph{Neal's Funnel.} As a concrete example, take Neal's funnel and the diagonal approximation to the Hessian
\begin{equation}
    C(V_n,X_n)=\begin{pmatrix}
        \left(\sigma^{-2}+\frac{1}{2}\exp(-v)\|x\|^2+\varepsilon\right)^{-1}&0\\
        0&\left(\exp(-v) + \varepsilon\right)^{-1}I
    \end{pmatrix},\notag
\end{equation}
for some $\varepsilon>0$, a regularization term to ensure positive definiteness. Figure~\ref{fig:nealsfunel} shows the samples of the Metropolis-adjusted, named FA-MALA, chain when $x\in\mathbb{R}^{10}$ using $20\,000$ burn-in iterations, and $50\,000$ samples taken at every $5$ iterations. It is clear that using this state-dependent preconditioning matrix allows the chain to sample deep into $v\ll 0$ while also enabling diffusive motion in $x_1$ when $v\gg 0$. This contrasts with MALA, which cannot sample deep into the neck and fails to effectively explore $x_1|v$ when $v\gg0$.

\subsection{State-dependent time rescaling}
Time rescaling can be seen as a specific version of position-dependent preconditioning, simplified to the scalar-mobility case $C(x)=r(x)I$, with $r(x)>0$:
\begin{equation}
    dX_t=\bigg(-r(X_t)\nabla U(X_t)+\nabla r(X_t)\bigg)\,dt + \sqrt{2r(X_t)}\,dW_t.
\end{equation}
Ideally, we would like to monitor the force magnitude $\|\nabla U(X_t)\|$ and rescale $r(X_t)$, the effective step size, to be small when the force is large and vice versa. This is not itself a curvature estimate, although in funnel examples it may correlate with local stiffness. As taken in \cite{Leroy2024}, one can consider $r(x)=\psi(\|\nabla U(x)\|)$, where
\begin{equation}
    \psi(s)=\frac{M\sqrt{1+m^2\kappa s^{2\alpha }}}{{\sqrt{1+m^2\kappa s^{2\alpha}}} + M\sqrt{\kappa s^{2\alpha}}},\notag
\end{equation}
where $\psi:\mathbb{R}_+\to[Mm/(m+M),M]$ is decreasing and $\kappa$ controls the decay rate. Larger $\kappa$ gives more rapid decay. In practice, since $r(x)$ involves the norm of $\nabla U(x)$, computing $\nabla r(x)$ can be computationally intensive. For continuous-time SDEs, a recent paper has shown how to perform this without requiring the additional $\nabla r(x)$ correction \cite{leimkuhler2025}.

For Metropolis-adjusted samplers, we instead treat the log step size $\ell=\log h$ as a random variable whose conditional mean decreases as $\|\nabla U(x)\|$ increases, and vice versa \cite{BouRabee2025, BouRabee2026, Biron2023, Grazzi2026}. Concretely, when the step size $h$ is fixed, the acceptance rate is
\begin{equation}
    \alpha(z,z')=\min\left(1,\frac{\widetilde{\pi}(z')q_h(z',z)}{\widetilde{\pi}(z)q_h(z,z')}\right),\notag
\end{equation}
where the $h$ subscript makes it clear that the proposal kernel depends on $h$. If instead $\ell\mid z\sim r_\ell(\cdot\mid z)$ and $h=e^\ell$, the Metropolis adjustment becomes
\begin{equation}
    \alpha(z,z')=\min\left(1,\frac{\widetilde{\pi}(z')q_h(z',z)}{\widetilde{\pi}(z)q_h(z,z')}\frac{r_\ell(\ell\mid z')}{r_\ell(\ell\mid z)}\right).
\end{equation}
For example, consider
\begin{equation}\label{eq:randomized_stepsize_probability}
    r_\ell(\ell\mid x)= \mathrm{TN}(\ell;\log \mu (x), \sigma^2, \log h_{\min}, \log h_{\max}),\quad \mu(x)=\frac{h_*}{\left(1+\left(\frac{\|\nabla U(x)\|}{\sqrt{d}}\right)^2\right)^{1/2}},
\end{equation}
where $\mathrm{TN}(\ell;\mu_0, \sigma_0^2, x_{\min}, x_{\max})$ denotes the density at $\ell$ of a truncated Normal distribution centered at $\mu_0$ with variance $\sigma_0^2$, truncated to $[x_{\min}, x_{\max}]$. This allows the sampler to respond probabilistically to regions of both small and large force magnitude. 

\begin{remark}
    Delayed-rejection schemes provide an alternative approach for sampling from funnel-like geometries \cite{Tierney1999, Mira2001, Modi_2024, turok2024}. However, they introduce branching instructions through sequential accept-reject decisions, which can complicate implementation and reduce the efficiency of vectorized computation. By contrast, time-rescaling methods retain a simpler computational structure and are straightforward to vectorize.
    
    Delayed rejection and NUTS respond to poor proposal scales in essentially opposite ways. A delayed-rejection method first attempts an aggressive proposal and, if it is rejected, switches to a more conservative kernel, typically by reducing the step size. NUTS instead composes many conservative integration steps to construct a longer trajectory, thereby producing large proposals without increasing the individual integration step size.
\end{remark}

\begin{algorithmidea}{Randomized Step Size OBABO MAKLA (RS-OBABO or RS-MAKLA)}
    Fix $h_*$ (tuned to target a fixed acceptance rate), and $\sigma=0.5$. Let $(x_n,p_n)$ be the current position and momentum of the chain.
    \begin{enumerate}
        \item Draw $\ell_n\mid x_n\sim r_\ell(\cdot\mid x_n)$ as defined in \eqref{eq:randomized_stepsize_probability}, and set $h_n=e^{\ell_n}$. The realized value $h_n$ is used in the OBABO discretization.
        \item Perform the first partial momentum refreshment $(\widetilde x_n, \widetilde p_n)=O(x_n,p_n)$,
        \begin{equation}
            \widetilde x_n = x_n,\quad  \widetilde p_n = \eta_n p_n+\sqrt{1-\eta^2_n}M^{1/2}\xi,\qquad \xi\sim\mathcal{N}(0, I),\notag
        \end{equation}
        where $\eta_n=e^{-\gamma h_n/2}$.
        \item Evolve according to the Hamiltonian flow BAB with time step $h_n$:
        \begin{equation}
            (x_n',p_n')=BAB(\widetilde x_n,\widetilde p_n).\notag
        \end{equation}
        \item Compute the Hamiltonian error accumulated during the BAB step,
        \begin{equation}
            \Delta=H(x_n',p_n')-H(\widetilde x_n,\widetilde p_n).\notag
        \end{equation}
        \item With probability
        \begin{equation}
            \alpha=\min\left(1,\exp(-\Delta)\frac{r_\ell(\ell_n\mid x_n')}{r_\ell(\ell_n\mid x_n)}\right),\notag
        \end{equation}
        set $(\overline x_n,\overline p_n)=(x_n',p_n')$; otherwise set
        $(\overline x_n,\overline p_n)=(\widetilde x_n,-\widetilde p_n)$.
        \item Perform the final partial momentum refreshment
        \begin{equation}
            (x_{n+1},p_{n+1})=O(\overline x_n,\overline p_n).\notag
        \end{equation}
    \end{enumerate}
    To improve performance, initialize $M$ to the Hessian at the mode and set the sampler's initial condition to the mode. During the burn-in phase, update the running covariance estimate $C_K$ and set $M_K=(C_K+\varepsilon I)^{-1}$.
    
    In practice, we instead use an OBABABO discretization \cite{chok2026} which gives higher-order accuracy. In this case, the Metropolis adjustment is applied to the Hamiltonian error accumulated during the BABAB step.
\end{algorithmidea}

\subsection{Why the combination matters}
No single mechanism resolves every geometric problem
\begin{itemize}
    \item NUTS adapts trajectory length, but a global leapfrog step size still has to resolve the narrowest region.
    \item A dense mass matrix removes global correlation but not exponential variation in conditional scale.
    \item A local time-rescaling mechanism can stabilize a funnel but may remain inefficient if the globally transformed coordinates are strongly correlated.
    \item A sophisticated local metric can be computationally counterproductive if its derivatives and linear algebra cost more than the saved gradient evaluations.
\end{itemize}
The strongest practical strategy is therefore layered:
\begin{enumerate}
    \item Parameterization: If possible, remove any funnel-like geometry from the distribution
    \item Global preconditioning: Precondition the dynamics through a Hessian at the mode (or MAP) computation before running the sampler
    \item Local scale adaptation: Utilize a dynamic step size so the sampler remains robust across regions with substantially different force magnitudes.
\end{enumerate}

\subsection{Which sampler works where?}
\paragraph{NUTS.} NUTS is often the strongest default for moderate-dimensional, differentiable Bayesian posteriors with full-data gradients. It offers a simple interface for practitioners and solves many of HMC's issues. One remark is to try dense-matrix preconditioning if the NUTS sampler doesn't work with diagonal preconditioning. 

NUTS is not automatically optimal. Tree building has irregular control flow, making it expensive to run on GPUs, and, at very high dimensions, storing and traversing large trees can be costly in both memory and time.

\paragraph{MALA.} MALA is most attractive when a local, fixed-cost proposal is desired and a global preconditioner captures most of the target geometry. Its local nature can, however, make it less efficient than persistent or trajectory-based methods when useful exploration requires long coherent movement.

In the case where a non-scalar position-dependent preconditioning may be beneficial over its scalar counterpart, MALA can prove to be useful. This, however, comes with the additional cost that (1) Vectorization of the kernels becomes trickier and not as efficient on GPUs, and (2) one must compute the Cholesky factorization of the preconditioning matrix.

% In optimization, MALA serves a different purpose and is mainly used for ...

\paragraph{MALT and MAKLA}. Underdamped methods serve as a midway between MALA and HMC. They offer the computational benefits of MALA, simple kernels that can be easily vectorized, use one gradient per iteration (for the OBABO discretization, and two gradients for OBABABO), and reduce random-walk behavior, as in HMC. Moreover, by using randomized step sizes, they offer the benefits of simple kernels and robustness to funnel-like geometry. 

\subsection{Time-rescaling and locally metric-aware methods}
Local adaptation is most valuable when the posterior contains genuine state-dependent scale variation after reasonable reparameterization and global whitening. While it may seem that local adaptation should always be applied, the additional term in the accept/reject step may reduce the acceptance rate, thereby requiring a smaller $h$ on average than a fixed-$h$ kernel.

Full dense matrix metric-aware methods are theoretically attractive; however, their additional complexity must be evaluated honestly. The local metric may require Hessians of the log-density and their Cholesky determinants, which add additional computational complexity. Moreover, its chain-dependent tree depths create irregular control flow, creating additional computational overhead.

\section{When not to use a Metropolis Correction}
For a dataset of size $N$, the log-posterior takes the form
\begin{equation}
    \log \pi(x) = \log p(x) + \sum_{i=1}^N\log p(y_i|x).\notag
\end{equation}
For the samplers discussed here, there are two main computationally intensive steps: (1) evaluating the posterior, and (2) evaluating the gradient. These issues become more apparent when the number of data points $N$ or the dimension of $x$ is large. In this case, adding a potential rejection during the Metropolis-correction stage can waste substantial computation. 

A primary example of such a problem is Bayesian neural networks, or Bayesian Kolmogorov--Arnold networks. These problems are high-dimensional, and expensive to evaluate on the full dataset. For the former case, recent papers have shown the effectiveness of unadjusted methods that use minibatch evaluations of the posterior (and, by extension, minibatch gradients) \cite{Paulin2025, Zuo2025, leimkuhler2025}.

\section{Discussion and Conclusion}
There is no universal winner. 
\begin{itemize}
    \item NUTS exploits long coherent paths, at the cost of a variable number of gradient evaluations. A binary tree of depth $D$ can contain up to order $2^D$ leapfrog steps: the computational cost is linear in the number of steps actually constructed, while the maximum permitted work grows exponentially with the maximum tree depth. 
    
    The No-U-Turn condition mimics the Exact HMC quarter-period sampler, which yields exact independent samples for the identity Gaussian case. As such, while NUTS yields excellent \textit{per-iteration} performance, it comes at the cost of additional gradient computations and computational time. However, when measuring the efficiency of the sampler in terms of effective sample size (ESS), NUTS often performs very well, as measured by ESS/gradient-evaluation or ESS/second. 

    NUTS' main disadvantage is that the proposal kernels incur additional computational overhead compared to fixed-work kernels. 
    
    \item RS-MAKLA maintains simple kernels (like MALA) while reducing random-walk behavior (like NUTS), and is most effective when GPUs are available, as these kernels can be vectorized efficiently and are extremely efficient on GPUs. Indeed, \cite{chok2026} exploits the vectorization of RS-MAKLA and outperforms NUTS on a wide range of posterior distributions. 
    % As such, when performing inference on Bayesian neural networks, in which GPUs are a must, MAKLA or MALA proposals are far more attractive than NUTS.
\end{itemize}

% There comes an interesting question on fair comparison. ESS/gradient-evaluation is, perhaps, the fairest implementation-independent comparison. However, one ultimately cares about the ESS/second. This raises an important question: Suppose we are limited to one CPU and have a small enough problem that a 5-chain vectorized implementation of MAKLA/MALA runs in about the same time as a 1-chain MAKLA/MALA implementation. NUTS, however, would have to run 5 chains sequentially. Typically, ESS/second is computed as ESS/(average time taken per chain). This, however, ignores the total runtime cost, for which MAKLA/MALA is 5 times more efficient due to its vectorization advantages. However, when run on a supercomputer with access to thousands of CPU cores, parallelizing NUTS is trivial. This, in turn, raises the question: since MAKLA/MALA can run 5-chains on each CPU core, should one run $5\times\mathrm{Number}\  \mathrm{of}\ \mathrm{Cores}$ to compute the ESS?

We recommend the following guideline for practitioners
\begin{enumerate}
    \item If the problem is (really) high-dimensional and the number of data points is large, resort to unadjusted underdamped Langevin methods (aka. OBABO, or OBABABO discretizations) with batch-gradient computations.
    \item Otherwise, use NUTS.
    \item If NUTS is slow, or the R-hat statistic is too large, first try an initial MAP preconditioning, and use a dense matrix preconditioning (see Appendix~\ref{app:RCode}).
    \item If NUTS still fails to produce good samples, the distribution often follows a hierarchical posterior that induces a funnel-like geometry. If possible, reparameterize the model to the non-centered form and run NUTS on it.
    \item If non-centering is not possible, or computationally impractical, use Randomized Step Size OBABO (or OBABABO) MAKLA. 
\end{enumerate}
For distributions with complex constraints, such as posteriors with a GEV likelihood, the issue can usually be resolved through an initial MAP preconditioning step. When more complex constraints arise (e.g., simplex or doubly-stochastic matrix constraints, which have zero Lebesgue measure), one must resort to more advanced sampling techniques.

\section*{Why Another Paper on Sampling?}
This paper was largely a result of interactions with my poster at the 2026 Center for Statistics Day at the University of Edinburgh. Through these interactions, I became aware of the substantial divide between, on the one hand, the theoretical analysis and design of new samplers and, on the other, the application of sampling methods in statistics.

I have often seen NUTS (or HMC) used as the default standard sampler in modern statistical applications. From my viewpoint, this is largely due to two reasons: (1) it is an exceptionally general-purpose sampler, and (2) it has an easy user interface for statisticians. To my knowledge, it is not common that any given sampler can outperform a well-tuned NUTS in terms of ESS/gradient or ESS/second. If it does, it rarely comes with the same level of software support and ease of use.\footnote{I am guilty of this.} Recent code-generation tools, including large language models, have reduced this implementation barrier, although they have not eliminated the need for careful validation.\footnote{If you have a problem for which NUTS is having difficulty sampling, feel free to email me.}

The main practical hurdle is often not how to replace NUTS, but how to make it work efficiently for the problem at hand. Substantial improvements can frequently be obtained by (1) using a dense preconditioning matrix, (2) an initial Hessian at MAP preconditioning, or (3) using non-centered parameterizations. These strategies are not universally beneficial, but they can dramatically improve performance when they reflect the posterior's geometry.

A more fundamental limitation arises when the target distribution is supported on a lower-dimensional constrained space, e.g., sets with zero Lebesgue measure. Examples include the probability simplex, doubly stochastic matrices, and manifolds of orthogonal matrices. Some such spaces, including the simplex, admit standard transformations to unconstrained coordinates. However, these transformations can distort the geometry of the target distribution, introduce difficult Jacobian terms, or become poorly conditioned near the boundary. For more complicated constraints, a convenient global parameterization may not exist at all. In such cases, one may need algorithms that operate directly on the constrained space and are designed specifically around its geometry.

\bibliographystyle{unsrt}

\newpage
\appendix
\section{Applications}
This appendix compares NUTS with the randomized step-size MAKLA (with an OBABABO discretization) on three centered hierarchical targets. 

\subsection{Common Sampling setup}
All experiments use ten independent chains, $5\,000$ warmup iterations per chain, and $10\,000$ retained iterations per chain, with no thinning. The German credit and radon experiments use exactly the same warmup and sampling lengths as Neal's funnel experiment. Unless explicitly labeled non-centered, both samplers target the same centered parameterization.

For NUTS, step-size and Euclidean metric adaptation are confined to warmup and are frozen before samples are retained. For RS-MAKLA, one OBABABO proposal is made per iteration. Conditional on the current position $x_n$, the log step size is drawn from \eqref{eq:randomized_stepsize_probability}, with $h_{\min}=10^{-4}$, $h_{\max}=1$, and log-scale standard deviation $0.5$. The momentum persistence is coupled to the realized step size through $\eta_n=\exp(-\gamma h_n/2)$, with $\gamma=0.1$. The reference scale $h_*$ and the global preconditioner are adapted only during warmup; the latter is initialized from local curvature at the mode and then updated as the regularized inverse of the running covariance. All adaptation is frozen for the retained iterations. In the Neal's funnel experiment, where no fixed global metric can remove the position-dependent scale, both methods instead use the identity metric.

We report effective sample size per gradient evaluation (ESS/Grad.) and per second (ESS/Sec.), together with rank-normalized split $\widehat R$. ESS/Grad. is the more implementation-independent measure of algorithmic efficiency. ESS/Sec. also reflects language, compiler, hardware, vectorization, and implementation maturity; it should therefore be interpreted as the realized performance of these implementations rather than an intrinsic ordering of the algorithms. The funnel plots pool retained draws across chains and are intended to diagnose geometric coverage. They do not replace trace plots, $\widehat R$, ESS, or divergence diagnostics.

\subsection{Neal's Funnel}
We consider \eqref{eq:neals_funnel} with $\sigma=3$ and $d=10$, giving an eleven-dimensional target after including $v$. For NUTS, we compared a fixed identity metric with adapted diagonal and dense Euclidean metrics. The identity metric gave the largest ESS and smallest $\widehat R$, and is therefore used below. This is consistent with the geometry: a fixed linear transformation cannot simultaneously whiten the narrow neck and the wide mouth. RS-MAKLA likewise uses $M=I$, so the comparison isolates adaptation to the local time scale rather than global preconditioning.

\begin{figure}
    \centering
    \includegraphics[width=0.9\linewidth]{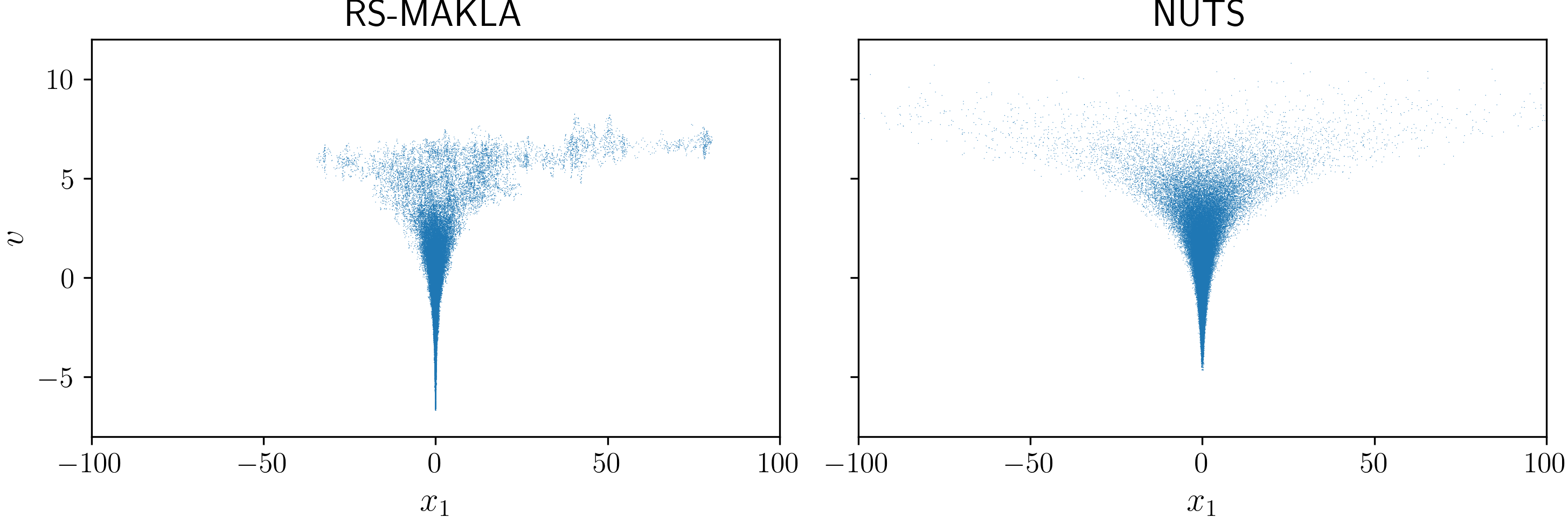}
    \caption{Retained samples from Neal's funnel projected onto a representative pair $(x_1,v)$. RS-MAKLA is labeled RS-OBABABO in the left panel. NUTS explores the broad mouth extensively, while the randomized-step-size chain reaches further into the narrow negative-$v$ region in this run. Samples are pooled across ten chains.}
    \label{fig:makla_neals_funnel}
\end{figure}

\begin{figure}
    \centering
    \includegraphics[width=0.5\linewidth]{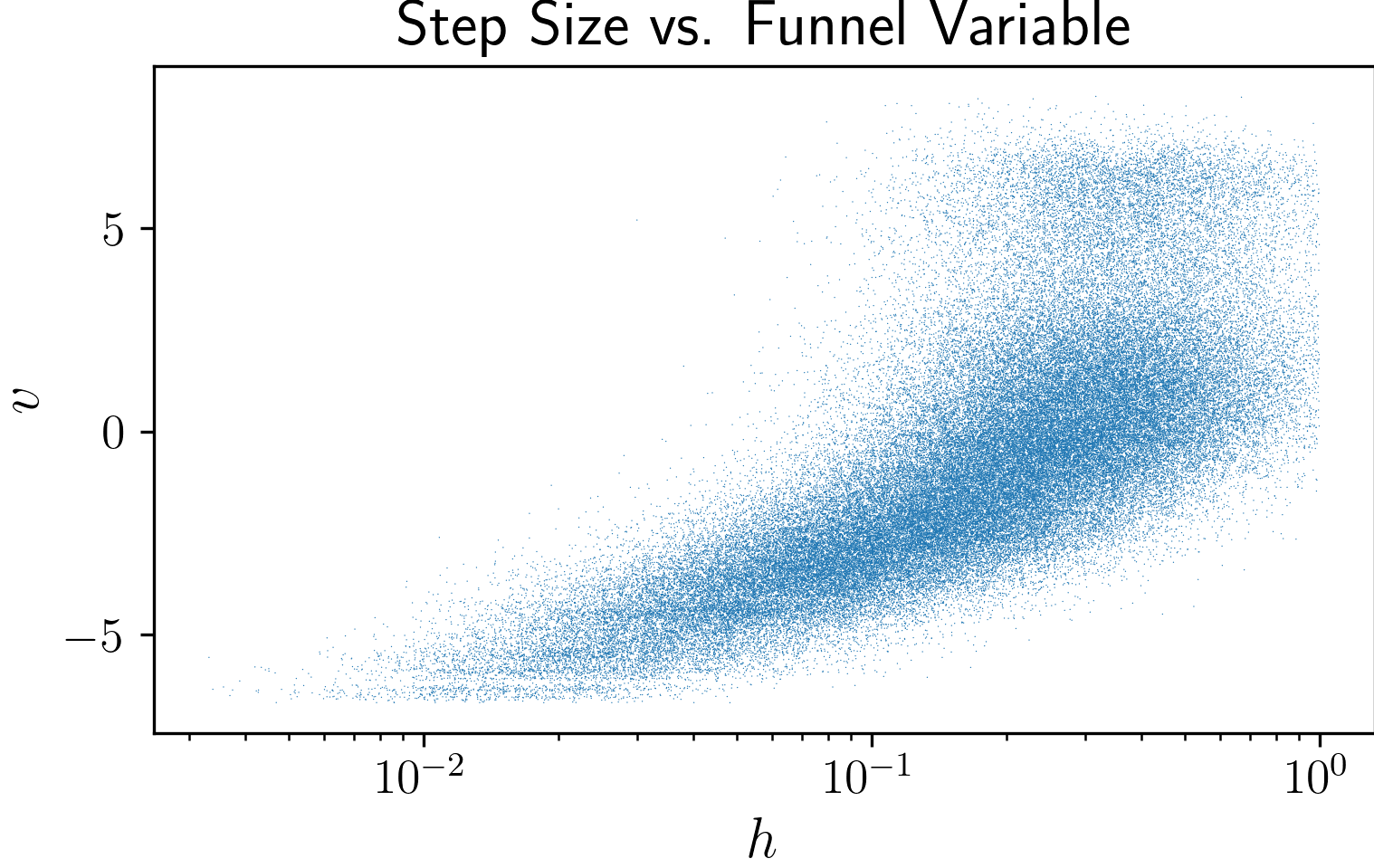}
    \caption{Realized RS-MAKLA step size $h$ against the funnel variable $v$ for Neal's funnel. The logarithmic horizontal axis makes the local time rescaling explicit: small steps are selected in the high-curvature neck and much larger steps in the low-curvature mouth.}
    \label{fig:makla_neals_stepsize}
\end{figure}

Figure~\ref{fig:makla_neals_funnel} shows that the two samplers respond differently to the strongest funnel geometry. NUTS explores the broad mouth effectively through long Hamiltonian trajectories, but its globally adapted leapfrog step size must remain stable in the narrowest region visited. RS-MAKLA reaches further into the negative-$v$ neck in this run while continuing to move across the mouth. Figure~\ref{fig:makla_neals_stepsize} explains the mechanism: the realized step size decreases by roughly two orders of magnitude between the wide and narrow parts of the funnel. The Metropolis correction, including the ratio of the conditional step-size densities, retains exactness while allowing the local numerical resolution to change with position.

This example is the regime in which RS-MAKLA is most attractive. A single scale variable controls ten conditionally Gaussian coordinates, so moving into the neck simultaneously increases the curvature of an entire block. NUTS can compensate by choosing a globally small step size and adapting trajectory length, whereas RS-MAKLA directly addresses the state-dependent stiffness. The latter has predictable, regular per-iteration work and can be easily vectorized across many chains, but it does not capture the long-distance movement of a NUTS trajectory from a single iteration.

\subsection{German Credit Dataset}
We next consider the hierarchical logistic regression model for the German credit dataset \cite{hofmann1994_data}, which is used as a difficult automatic reparameterization example in \cite{Gorinova_2020}. Writing $\rho_i=\log\tau_i$, the centered hierarchy is
\begin{align}
    \rho_0 &\sim \mathcal{N}(0,10^2),
    &\rho_i\mid\rho_0 &\sim \mathcal{N}(\rho_0,1), \\
    \beta_i\mid\rho_i &\sim \mathcal{N}(0,e^{2\rho_i}),
    &y_j\mid\beta &\sim \operatorname{Bernoulli}\!\left(\operatorname{logit}^{-1}(X_j^\top\beta)\right).
\end{align}
Here each local scale $\tau_i$ controls only one coefficient $\beta_i$, while $\rho_0$ couples all local scales. The geometry is therefore a collection of low-dimensional local funnels, with an additional weaker global dependence.

\begin{figure}
    \centering
    \includegraphics[width=0.9\linewidth]{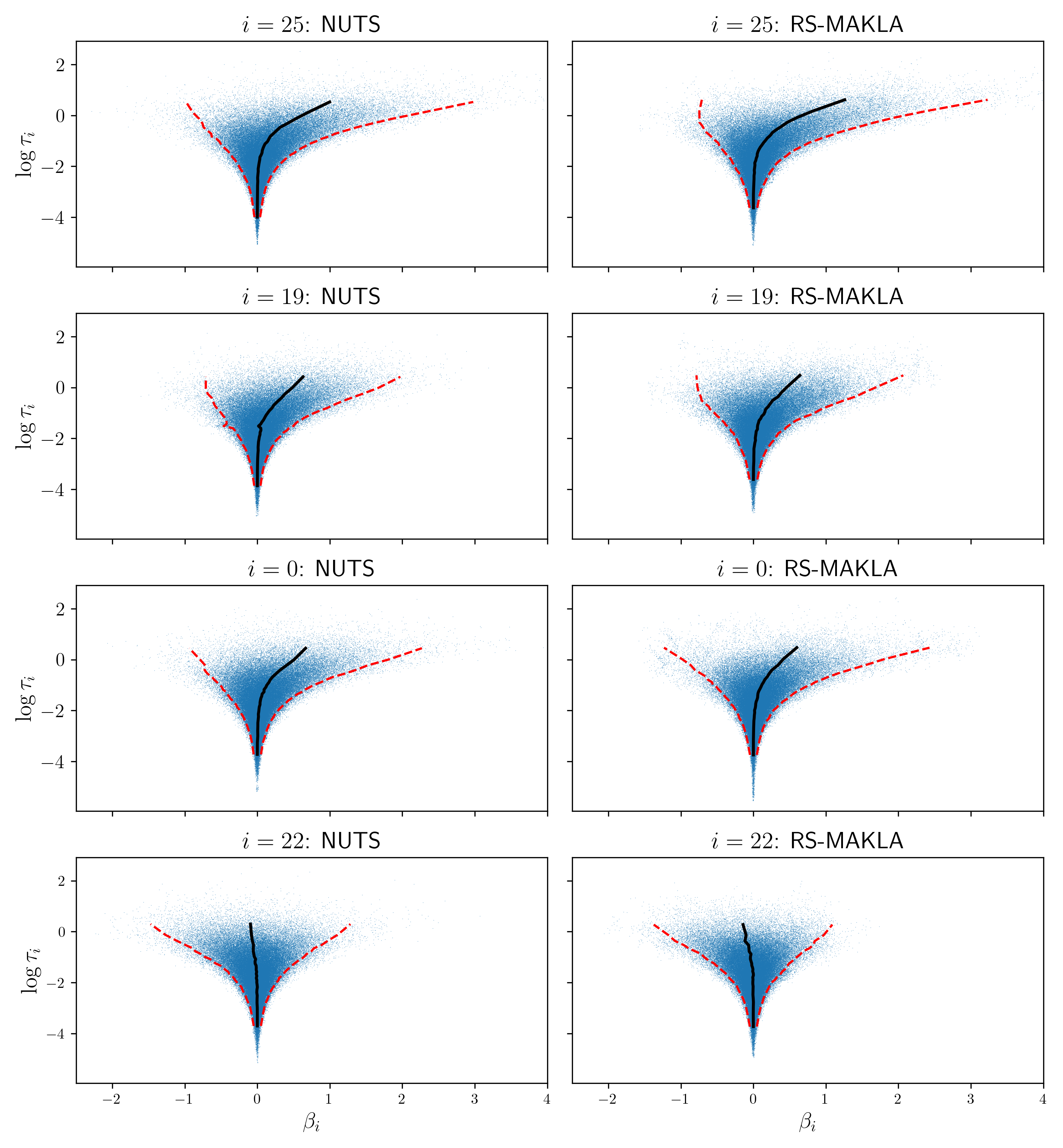}
    \caption{Centered German credit posterior for four representative local scale--coefficient pairs. The black solid curve is the conditional center and the red dashed curves show the conditional $\pm2$ standard-deviation envelope. Both methods enter the local necks, while NUTS gives the stronger aggregate diagnostics in Table~\ref{tab:german_diagnostics}. Samples are pooled across chains.}
    \label{fig:german_funnel}
\end{figure}

Figure~\ref{fig:german_funnel} shows that both methods reproduce the conditional funnel shape and visit its narrow region. Two features make this target substantially easier for NUTS than Neal's funnel. First, the likelihood suppresses the most extreme values of the local scales. Second, entering the neck associated with one $\tau_i$ stiffens essentially one coefficient direction, rather than all coefficient directions at once. A global NUTS step size is consequently less severely constrained, while its long trajectories efficiently traverse the correlated posterior.

The quantitative comparison in Table~\ref{tab:german_diagnostics} favors NUTS overall. Its median ESS/Grad. is about $2.5$ times larger for the coefficient block and $2.1$ times larger for the log-scale block. The exception is the worst log-scale coordinate: RS-MAKLA has a larger minimum ESS/Grad., indicating that local step-size reduction can protect the most difficult scale direction even when it does not improve the median. NUTS also has uniformly smaller $\widehat R$. RS-MAKLA has thirteen coordinates slightly above the strict $1.01$ threshold, with a maximum of $1.012$; this is not a severe failure, but it indicates that the retained run is less uniformly mixed than the NUTS run. The much larger ESS/Sec. values for NUTS show that, for this CPU implementation and this moderate-dimensional model, regular one-step proposals do not offset the benefit of longer NUTS trajectories.\footnote{ESS/Sec. was computed as ESS/(average time per chain), not total runtime.}

\begin{table}
    \centering
    \begin{tabular}{l|rr|rr}
    \toprule
    Sampler & \multicolumn{2}{c|}{RS-MAKLA} & \multicolumn{2}{c}{NUTS} \\
     & $\beta$ & $\log\tau_i$ & $\beta$ & $\log\tau_i$ \\
    \midrule
    $\min \mathrm{ESS}/\mathrm{Grad.}$ & 8.66e-03 & 7.74e-03 & 1.15e-02 & 3.26e-03 \\
    $\mathrm{med}\,\mathrm{ESS}/\mathrm{Grad.}$ & 1.79e-02 & 1.39e-02 & 4.47e-02 & 2.93e-02 \\
    $\max \mathrm{ESS}/\mathrm{Grad.}$ & 7.57e-02 & 2.00e-02 & 9.97e-02 & 9.99e-02 \\
    \midrule
    $\min \mathrm{ESS}/\mathrm{Sec.}$ & 1.48e+01 & 1.33e+01 & 2.73e+02 & 7.76e+01 \\
    $\mathrm{med}\,\mathrm{ESS}/\mathrm{Sec.}$ & 3.07e+01 & 2.37e+01 & 1.07e+03 & 6.97e+02 \\
    $\max \mathrm{ESS}/\mathrm{Sec.}$ & 1.30e+02 & 3.43e+01 & 2.37e+03 & 2.38e+03 \\
    \midrule
    $\min \hat{R}$ & 1.001 & 1.003 & 1.000 & 1.000 \\
    $\mathrm{med}\,\hat{R}$ & 1.005 & 1.007 & 1.001 & 1.001 \\
    $\max \hat{R}$ & 1.011 & 1.012 & 1.002 & 1.004 \\
    $\#(\hat{R} > 1.01)$ & 5 & 8 & 0 & 0 \\
    \bottomrule
    \end{tabular}
    \caption{German credit diagnostics over the coefficient block $\beta$ and log-scale block $\rho=(\rho_0,\rho_1,\ldots)$. ESS/Sec. is implementation- and hardware-dependent; ESS/Grad. is the cleaner comparison of proposal efficiency.}
    \label{tab:german_diagnostics}
\end{table}

\subsection{Minnesota radon data}
For the Minnesota radon data \cite{Gelman2006}, let $c[i]$ denote the county of observation $i$, let $x_i$ be the floor indicator, and let $u_c$ be a county-level uranium covariate. We use the centered varying-intercept model
\begin{align}
    &\mu,a,b\sim\mathcal{N}(0,1),\quad \log \tau\sim \mathcal{N}(0,1),\quad \log\sigma\sim\mathcal{N}(0, 1),\notag\\
    &m_c|\mu,a,\tau \sim\mathcal{N}(\mu + au_c,\tau^2),\quad \log(r_i + 0.1)|m,b,\sigma \sim\mathcal{N}(m_{c[i]} + bx_i, \sigma^2).\notag
\end{align}
Unlike the German credit hierarchy, the single scale $\tau$ controls all $85$ county effects. Structurally, this is much closer to Neal's funnel: when $\tau$ is small, an entire block of county effects contracts simultaneously.

\begin{figure}
    \centering
    \includegraphics[width=0.9\linewidth]{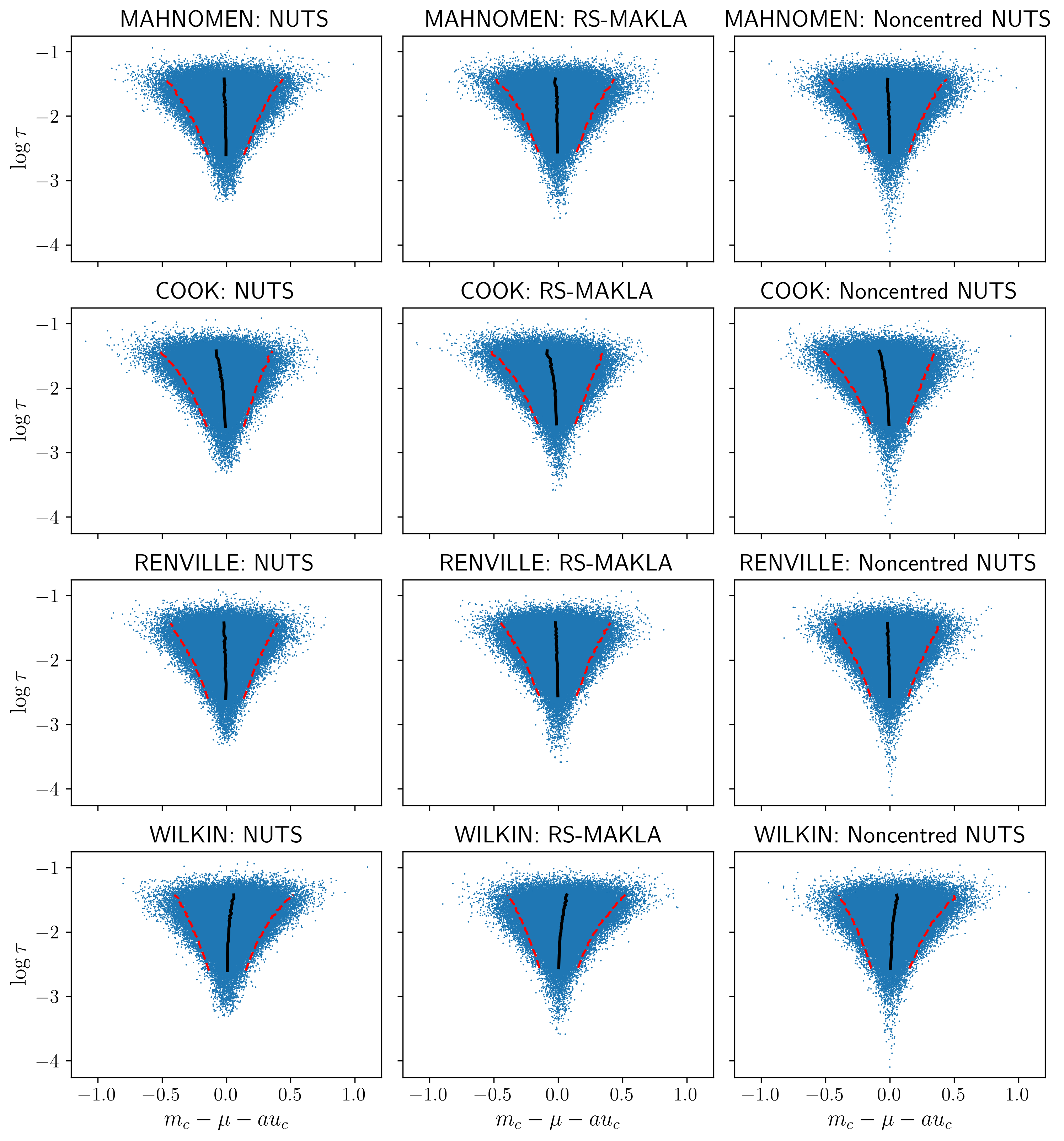}
    \caption{Minnesota radon posterior for four representative counties. The first two columns show centered NUTS and centered RS-MAKLA; the third shows non-centered NUTS transformed back to $(m_c-\mu-a u_c,\log\tau)$. The black curve is the conditional center and the red dashed curves show the conditional $\pm2$ standard-deviation envelope. Samples are pooled across chains.}
    \label{fig:radon_funnel}
\end{figure}

The important distinction is that the data prevent the posterior from exploring the most extreme part of the prior funnel. In Figure~\ref{fig:radon_funnel}, the sampled log scale is concentrated in a comparatively short interval, so the curvature variation encountered by the chains is far smaller than in Neal's funnel. Consequently, centered NUTS remains highly effective. It has a median ESS/Grad. about $5.2$ times larger for the county-effect block than RS-MAKLA, while the two methods have almost identical ESS/Grad. for the scalar log scale. This separation is informative: the randomized local step size robustly handles the difficult scale coordinate, but a one-step underdamped proposal does not match the long, coherent NUTS trajectories for moving through the $85$-dimensional effect block. Both runs have excellent $\widehat R$, so the difference is efficiency rather than evident nonconvergence.

The right column of Figure~\ref{fig:radon_funnel} shows NUTS applied to the non-centered representation $m_c=\mu+a u_c+\tau z_c$, $z_c\sim\mathcal{N}(0,1)$, transformed back to the centered coordinates for plotting. Non-centering is therefore a natural first alternative to test when the centered hierarchy exhibits funnel-like behavior. In this data-rich example, however, the centered posterior is already sufficiently regular that centered NUTS performs well. The figure therefore illustrates an important practical point: the existence of a hierarchical scale controlling many parameters signals potential funnel geometry, but the realized posterior difficulty depends on how strongly the likelihood truncates the neck.

\begin{table}
    \centering
\begin{tabular}{l|rr|rr}
\toprule
Sampler & \multicolumn{2}{c|}{RS-MAKLA} & \multicolumn{2}{c}{NUTS} \\
& $m$ & $\log\tau$ & $m$ & $\log\tau$ \\
\midrule
$\min \mathrm{ESS}/\mathrm{Grad.}$ & 6.24e-02 & 3.11e-02 & 1.32e-01 & 3.07e-02 \\
$\mathrm{med}\,\mathrm{ESS}/\mathrm{Grad.}$ & 1.23e-01 & 3.11e-02 & 6.42e-01 & 3.07e-02 \\
$\max \mathrm{ESS}/\mathrm{Grad.}$ & 2.40e-01 & 3.11e-02 & 8.26e-01 & 3.07e-02 \\
\midrule
$\min \mathrm{ESS}/\mathrm{Sec.}$ & 2.69e+02 & 1.34e+02 & 5.38e+03 & 1.25e+03 \\
$\mathrm{med}\,\mathrm{ESS}/\mathrm{Sec.}$ & 5.29e+02 & 1.34e+02 & 2.63e+04 & 1.25e+03 \\
$\max \mathrm{ESS}/\mathrm{Sec.}$ & 1.03e+03 & 1.34e+02 & 3.38e+04 & 1.25e+03 \\
\midrule
$\min \hat{R}$ & 1.000 & 1.002 & 1.000 & 1.001 \\
$\mathrm{med}\,\hat{R}$ & 1.001 & 1.002 & 1.000 & 1.001 \\
$\max \hat{R}$ & 1.001 & 1.002 & 1.000 & 1.001 \\
$\#(\hat{R} > 1.01)$ & 0 & 0 & 0 & 0 \\
\bottomrule
\end{tabular}
    \caption{Minnesota radon diagnostics for the county-effect block $m=(m_1,\ldots,m_{85})$ and the hierarchical log scale $\rho=\log\tau$. Both samplers mix well; NUTS is substantially more efficient for the high-dimensional effect block.}
    \label{tab:radon_diagnostics}
\end{table}

\subsection{Practical interpretation}
\label{app:application_summary}

The three examples separate prior funnel structure from posterior computational difficulty. Neal's funnel retains an extreme neck in which one scale controls a full block, and local time rescaling allows RS-MAKLA to reach further into the negative-$v$ neck in this experiment. The German credit likelihood regularizes a collection of one-coordinate local funnels, so NUTS can exploit long trajectories without being dictated by a single globally stiff block. The radon model has a stronger one-to-many hierarchy, but the likelihood constrains the scale sufficiently that centered NUTS remains highly efficient.

These results suggest the following ordering of interventions.
\begin{enumerate}
    \item Reparameterize to a non-centered form when available and when it improves the posterior geometry. 
    \item Otherwise, use NUTS as the default for moderate-dimensional CPU-based inference, especially when warmup yields a reliable Euclidean metric, and the posterior does not exhibit an extreme neck.
    \item RS-MAKLA is most compelling when the posterior remains strongly multiscale after global preconditioning, if reparameterization is computationally impractical, when predictable one-step kernels are needed, or when many chains can be vectorized on accelerators. Indeed, \cite{chok2026} exploits the vectorization of RS-MAKLA and outperforms NUTS on a wide range of posterior distributions. 
\end{enumerate}

\section{Identity Gaussian Underdamped Langevin}\label{sec:id_gaus_underdamped_langevin}
With $M=I$, define $Z_t=(X_t,P_t)$. Then
\begin{equation}
    dZ_t=AZ_t\,dt+B\,dW_t,\quad A=\begin{pmatrix}
        0&I\\ -I&-\gamma I
    \end{pmatrix},\quad B=\begin{pmatrix}
        0\\\sqrt{2\gamma}I
    \end{pmatrix}.\notag
\end{equation}
The transition is Gaussian:
\begin{equation}
    Z_t=e^{At}Z_0+\zeta_t,\quad \zeta_t\sim\mathcal{N}\left(0,\int^t_0 e^{As}BB^\top e^{A^\top s}ds\right).\notag
\end{equation}
For $0<\gamma<2$, the deterministic dynamics are damped oscillations with
frequency $\omega=\sqrt{1-\gamma^2/4}$. Unlike overdamped Langevin, whose position correlations decay monotonically, underdamped correlations may cross zero. As such, momentum can reduce diffusive behavior by carrying the process through the typical set. Small $\gamma$ preserves long oscillatory memory, whereas increasing $\gamma$ damps the motion and, after an appropriate time rescaling, moves the position process toward the overdamped regime.

\section{R Code -- Preconditioning at MAP}\label{app:RCode}
\begin{minted}{R}
nuts_from_map <- function(stan_file, data = list(), chains = 4, ...) {
  mod <- cmdstan_model(stan_file, compile_model_methods = TRUE)

  # MAP of the density used by NUTS
  map <- mod$optimize(data = data, jacobian = TRUE, seed = 0)

  # Extract the MAP in Stan's unconstrained parameter space
  u_draws <- map$unconstrain_draws(format = "draws_matrix")
  u_map <- as.numeric(u_draws[[1]])

  # Hessian of the unconstrained log posterior at the MAP
  H <- map$hessian(unconstrained_variables = u_map, jacobian = TRUE)$hessian

  H <- 0.5 * (H + t(H))  # Force H to be symmetric up to numerical precision

  # Posterior precision approximation
  precision <- -H

  # Ensure positive definiteness
  eig <- eigen(precision, symmetric = TRUE)
  eig$values <- pmax(eig$values, 1e-8)

  precision <- eig$vectors %*% diag(eig$values, nrow = length(eig$values)) %*% t(eig$vectors)

  # CmdStan's inv_metric is the approximate posterior covariance
  inv_metric <- solve(precision)
  inv_metric <- 0.5 * (inv_metric + t(inv_metric))

  fit <- mod$sample(data = data, init = map, chains = chains, 
                    parallel_chains = chains, metric = "dense_e", 
                    inv_metric = inv_metric, adapt_engaged = TRUE, ...)

  list(fit = fit, map = map, unconstrained_map = u_map, hessian = H,
       precision = precision, inv_metric = inv_metric)
}
\end{minted}

\section{Python Code -- Gaussian Laboratory}\label{app:gaussian_lab_python_code}
We give minimal code to run exact HMC, underdamped and overdamped Langevin code for this Gaussian setting. 
\begin{minted}{python}
import numpy as np

def exact_hmc(x, delta_t, rng=None):
    rng = np.random.default_rng() if rng is None else rng
    
    c, s = np.cos(delta_t), np.sin(delta_t)
    p = rng.normal(size=x.shape)
    return c * x + s * p, -s * x + c * p

def exact_underdamped(x, p, delta_t, gamma=1.0, rng=None):
    rng = np.random.default_rng() if rng is None else rng

    omega = np.sqrt(1.0 - gamma**2 / 4.0)
    c, s = np.cos(omega * delta_t), np.sin(omega * delta_t) / omega

    F = np.exp(-gamma * delta_t / 2.0) * np.array([
        [c + gamma * s / 2.0,  s],
        [-s, c - gamma * s / 2.0],
    ])

    noise_cov = np.eye(2) - F @ F.T
    noise = rng.multivariate_normal(np.zeros(2), noise_cov, size=x.size)

    x_next = F[0, 0] * x + F[0, 1] * p + noise[:, 0]
    p_next = F[1, 0] * x + F[1, 1] * p + noise[:, 1]

    return x_next, p_next


def exact_overdamped(x, delta_t, rng=None):
    rng = np.random.default_rng() if rng is None else rng

    a = np.exp(-delta_t)
    return a * x + np.sqrt(1.0 - a**2) * rng.normal(size=x.shape)
\end{minted}

\subsection{OBABO and OBABABO Discretizations}\label{app:obabo_obababo}
\begin{minted}{Python}
import numpy as np

def obabo(x, p, h, gamma=1.0, rng=None):
    rng = np.random.default_rng() if rng is None else rng
    eta = np.exp(-gamma * h / 2)

    p = eta * p + np.sqrt(1 - eta**2) * rng.normal(size=p.shape)
    p = p - 0.5 * h * x
    x = x + h * p
    p = p - 0.5 * h * x
    p = eta * p + np.sqrt(1 - eta**2) * rng.normal(size=p.shape)

    return x, p


def obababo(x, p, h, gamma=1.0, rng=None):
    rng = np.random.default_rng() if rng is None else rng

    b1 = (3 - np.sqrt(3)) / 6
    b2 = 1 - 2 * b1
    eta = np.exp(-gamma * h / 2)

    p = eta * p + np.sqrt(1 - eta**2) * rng.normal(size=p.shape)
    p = p - b1 * h * x
    x = x + 0.5 * h * p
    p = p - b2 * h * x
    x = x + 0.5 * h * p
    p = p - b1 * h * x
    p = eta * p + np.sqrt(1 - eta**2) * rng.normal(size=p.shape)

    return x, p


def run_underdamped(method, d, h, num_iters, gamma=1.0):
    thin = int(1 / h)
    history = np.zeros((num_iters + 1, d))

    x = np.zeros(d)
    p = np.zeros(d)
    rng = np.random.default_rng(1)

    for i in range(num_iters):
        for _ in range(thin):
            x, p = method(x, p, h, gamma=gamma, rng=rng)

        history[i + 1] = x

    return history


obabo_history = run_underdamped(obabo, d=10, h=0.01, num_iters=1000)
obababo_history = run_underdamped(obababo, d=10, h=0.01, num_iters=1000)
\end{minted}

\end{document}